\documentclass[a4paper]{report}
\usepackage[utf8]{inputenc}
\usepackage[T1]{fontenc}
\usepackage{RJournal}
\usepackage{amsmath,amssymb,array}
\usepackage{booktabs}

\providecommand{\tightlist}{%
  \setlength{\itemsep}{0pt}\setlength{\parskip}{0pt}}

\usepackage{longtable}

\makeatletter
 \let\@cite@ofmt\@firstofone
 \def\@biblabel#1{}
 \def\@cite#1#2{{#1\if@tempswa , #2\fi}}
\makeatother
\newlength{\cslhangindent}
\newlength{\csllabelwidth}
 {\begin{list}{}{%
  \setlength{\itemindent}{0pt}
  \setlength{\leftmargin}{0pt}
  \setlength{\parsep}{0pt}
  \ifodd #1
   \setlength{\leftmargin}{\cslhangindent}
   \setlength{\itemindent}{-1\cslhangindent}
  \fi
  \setlength{\itemsep}{#2\baselineskip}}}
 {\end{list}}
\usepackage{calc}

\usepackage[table]{xcolor}
\usepackage{float}
\usepackage{etoolbox}
\makeatletter
\ifpatchable{\@footnotetext}{\setspace@singlespace}{%
  \patchcmd{\@footnotetext}
    {\ignorespaces}
    {\Hy@raisedlink{%
       \hyper@anchorstart{\Hy@footnote@currentHref}\hyper@anchorend}%
     \ignorespaces}
    {}{\PackageError{dapper-paper}{Could not restore footnote hyperlinks}{}}
}{}
\makeatother

\begin{document}

\sectionhead{Contributed research article}
\volume{XX}
\volnumber{YY}
\year{20ZZ}
\month{AAAA}

\begin{article}
\title{dapper: Data Augmentation for Private Posterior Estimation in R}

\author{by Kevin Eng, Jordan A. Awan, Nianqiao Phyllis Ju, Vinayak A. Rao, and Ruobin Gong}

\maketitle

\abstract{%
This paper serves as a reference and introduction to using the R package dapper. dapper encodes a sampling framework which allows exact Markov chain Monte Carlo simulation of parameters and latent variables in a statistical model given privatized data. The goal of this package is to fill an urgent need by providing applied researchers with a flexible tool to perform valid Bayesian inference on data protected by differential privacy, allowing them to properly account for the noise introduced for privacy protection in their statistical analysis. dapper offers a significant step forward in providing general-purpose statistical inference tools for privatized data.
}

\section{Introduction}\label{sec:intro}

Differential privacy (DP) provides a rigorous framework for protecting
confidential information from re-identification attacks by using random
noise to obscure the connection between the individual and their data \citep{Dwork2006}.
Its development was spurred on by successful attacks
on anonymized data sets containing sensitive personal information. Prior
to differential privacy, anonymization schemes did not always have sound
theoretical guarantees despite appearing adequate.
Differential privacy marks a leap forward in the science of privacy by placing
privacy guarantees on a rigorous theoretical footing and moving beyond earlier
ad hoc approaches.
Several recent high-profile
implementations of differential privacy include those by Apple \citep{tang2017privacy}, Google \citep{Erlingsson_2014}, Microsoft \citep{ding2017collecting}, and the
U.S. Census Bureau \citep{Abowd2018}.

Many data sets amenable to differential privacy contain
valuable information that stakeholders are still interested in
learning about. However, the noise introduced by differential privacy
changes the calculus of inference. As an example,
we can implement differential privacy for tabular data by directly
adding independent, random error to each cell; the amount and type of which are determined by the DP mechanism design.
When we fit a regression model to the noise-infused data,
this will correspond to having measurement errors
in the covariates. This, unfortunately, violates the assumptions of most statistical models.
In the presence of such errors, standard estimators can exhibit significant bias and incorrect uncertainty quantification
\citep{Gong2022, karwa2015private, Wang2018}.
These issues are a serious concern for researchers \citep{SantosLozada2020, Kenny2021, Winkler2021}.
Given the importance of statistical methods in scientific research,
the increased adoption of privacy tools makes developing privacy-aware statistical workflows crucial.

Unfortunately, making the necessary adjustments poses formidable mathematical
challenges \citep{NIPS2010_sherry}, even for seemingly simple models like linear regression.
The difficulty lies in the marginal likelihood for the privacy-protected data.
This function is often analytically intractable and as a result,
it is difficult or impossible to apply traditional statistical methods
to derive estimators. In particular, the marginal likelihood can involve a complex
integral where it is not even possible to evaluate the likelihood
at a point. Tackling the problem by approximating the likelihood can be computationally
infeasible since the integral is usually high dimensional.
Few tools are available to researchers to address these issues,
and their absence is a serious barrier to the wider adoption
of differential privacy.

The \CRANpkg{dapper} package, short for Data Augmentation for Private Posterior Estimation in R, provides a set of tools for conducting
privacy-aware Bayesian inference. It serves as an R
interface for the data augmentation MCMC framework proposed by \citet{Ju2022},
allowing existing Bayesian models to be extended to handle
noise-infused data. The package is designed to integrate well with existing Bayesian workflows;
results can be analyzed using tools from the \CRANpkg{rstan} \citep{rstan_pkg} ecosystem in a drop-in fashion. Additionally,
construction of a privacy-aware sampler is simplified
through the specification of four independent modules: \texttt{latent\_f()}, \texttt{posterior\_f()}, \texttt{statistic\_f()},
and \texttt{mechanism\_f()}, which are described in detail later in the paper.
The benefits are twofold: several privacy mechanisms --- these can even be from different formal privacy frameworks ---
can be compared easily by only swapping out relevant modules. Furthermore, privacy mechanisms
that have non-smooth transformations resulting in the aforementioned intractable likelihoods
(see example 3 which involves clamping) can be incorporated with little work. As a result,
\CRANpkg{dapper} may prove particularly useful to those engaged in studying the privacy-utility trade-off or dealing with a privacy mechanism that involves multiple transformations.

\subsection{Related software and paper outline}\label{sec:related-software}

Software for differential privacy spans two distinct tasks: constructing
privacy-preserving releases and performing valid inference after a private
release has been observed. Table \ref{tab:software-comparison} compares
\CRANpkg{dapper} with the R packages \CRANpkg{tidydp} \citep{tidydp_pkg},
\CRANpkg{DPpack} \citep{DPpack_pkg}, \CRANpkg{diffpriv} \citep{diffpriv_pkg}, and
\CRANpkg{SimBaRepro} \citep{SimBaRepro_pkg}; the multilingual OpenDP library
\citep{opendp_pkg}; and the Python packages \texttt{d3p} \citep{d3p_pkg} and \texttt{diffprivlib}
\citep{diffprivlib_pkg}.

\begin{table}[!h]
\centering
\caption{\label{tab:software-comparison}Comparison of representative differential-privacy software. Rows are ordered by decreasing similarity to \CRANpkg{dapper}, prioritizing support for inference from an existing private release, followed by inferential paradigm and scope. The "Existing private release?" column indicates whether a tool performs inference conditional on an already released privatized statistic or data set; "Limited" denotes support restricted to a specialized setting. VI denotes variational inference. Among the tools compared, only dapper performs Bayesian inference from existing private releases with user-defined models and privacy mechanisms.}
\centering
\resizebox{\ifdim\width>\linewidth\linewidth\else\width\fi}{!}{
\fontsize{7}{9}\selectfont
\begin{tabular}[t]{>{\raggedright\arraybackslash}p{1.7cm}>{\raggedright\arraybackslash}p{2.2cm}>{\raggedright\arraybackslash}p{3.5cm}>{\raggedright\arraybackslash}p{2.4cm}>{\raggedright\arraybackslash}p{5.4cm}}
\toprule
Software & Language & Role and paradigm & "Existing private release?" & Scope and computation\\
\midrule
\cellcolor{gray!10}{dapper} & \cellcolor{gray!10}{R} & \cellcolor{gray!10}{Inference; Bayesian MCMC} & \cellcolor{gray!10}{Yes} & \cellcolor{gray!10}{User-defined models and mechanisms; latent-data MCMC}\\
SimBaRepro & R & Inference; frequentist & Yes & Simulation-based tests and confidence sets; no point estimate\\
\cellcolor{gray!10}{OpenDP} & \cellcolor{gray!10}{R, Python, Rust} & \cellcolor{gray!10}{Private release framework; specialized inference} & \cellcolor{gray!10}{Limited} & \cellcolor{gray!10}{Composable mechanisms; specialized binomial inference}\\
d3p & Python & Private fitting; Bayesian VI & No & NumPyro models; minibatched JAX on CPU/GPU\\
\cellcolor{gray!10}{DPpack} & \cellcolor{gray!10}{R} & \cellcolor{gray!10}{Private release and fitting; frequentist} & \cellcolor{gray!10}{No} & \cellcolor{gray!10}{Statistics, regression, and classification; model-specific}\\
diffprivlib & Python & Private release and fitting; frequentist & No & Mechanisms, accounting, and scikit-learn-style estimators\\
\cellcolor{gray!10}{diffpriv} & \cellcolor{gray!10}{R} & \cellcolor{gray!10}{Private release; not inferential} & \cellcolor{gray!10}{No} & \cellcolor{gray!10}{Generic mechanisms and sensitivity sampling; target-dependent}\\
tidydp & R & Private release; not inferential & No & Tidy counts, sums, and means; in-memory data frames\\
\bottomrule
\end{tabular}}
\end{table}

The closest inferential comparison is \CRANpkg{SimBaRepro} \citep{SimBaRepro_pkg},
an R implementation of the simulation-based, finite-sample inference framework
developed by \citet{Awan2025}. It also operates on an existing private release when the
privacy mechanism is known, but it provides finite-sample frequentist tests,
confidence intervals, and confidence sets rather than a point estimate or
posterior draws. The \texttt{d3p} package also
provides Bayesian methodology, but makes model fitting differentially private
rather than conditioning on a previously released private statistic. The
remaining tools primarily construct private releases or fit models directly to
confidential data under a privacy guarantee. OpenDP also provides specialized
inference utilities, such as confidence intervals and hypothesis tests for
binomial statistics released with its canonical noise mechanism, but
it does not provide general-purpose inference from arbitrary privatized statistics.

The rest of this article is organized as follows:
Section \hyperref[sec:background]{2} covers the necessary background to understand the mathematical notation
and ideas used throughout the paper. Section \hyperref[sec:methodology]{3} goes over the main algorithm without
going into mathematical detail; for specifics, see \citet{Ju2022}. Section \hyperref[sec:structure]{4} provides
an overview of the \CRANpkg{dapper} package and discusses important implementation details.
Section \hyperref[sec:examples]{5} contains four examples of how one might use the package to analyze the
impact of adding noise for privacy. The first example goes over a typical
odds ratio analysis for a \(2 \times 2\) table, the second example highlights
the modular nature of \CRANpkg{dapper} and reanalyzes the first example under a different
privacy scheme, the third example covers a linear regression model, and the fourth example
shows how \CRANpkg{dapper} can be used as a wrapper around an existing sampler. Finally, Section \hyperref[sec:computation]{6}
reports empirical performance for the examples and discusses an important practical implication of using a small
privacy loss budget with \CRANpkg{dapper}.

\section{Background}\label{sec:background}

Let \(x = (x_1, \ldots, x_n) \in \mathcal{X}^n\) represent a confidential
database containing \(n\) records. We assume that the records are independent and
identically distributed (i.i.d.) draws from a statistical model
\(f(\cdot \mid \theta)\), though this assumption can be relaxed. In many studies,
scientists are interested in learning about \(\theta\) because it provides important information about the scientific question of interest.

In the Bayesian statistical framework, learning about \(\theta\) is
accomplished by using the data \(x\) to update the
posterior \(p(\theta \mid x) \propto f(x \mid \theta) p(\theta)\).
Here, \(p(\theta)\) is called the prior distribution, and represents
the researcher's belief about \(\theta\) before seeing any data. The
posterior represents uncertainty around \(\theta\) and is formed by using Bayes' rule to fuse together the
observed data and the researcher's prior belief.
One major advantage of the Bayesian method is that, through the prior,
it provides a mechanism for incorporating information not explicitly contained
in the data at hand. This is especially useful in settings where there
is considerable domain knowledge about the value of \(\theta\).

When they are available, it is desirable to work with a summary statistic \(s = s(x)\)
that has much smaller dimension than the original data because doing so can
greatly simplify calculations. Since summary statistics are easier to work with,
database curators often publish them to efficiently communicate information
contained in large data sets.
This makes them a natural target for
dissemination-based privacy approaches \citep{Karr2016}.

\subsection{Differential privacy}\label{sec:dp}

While a summary statistic can already partially anonymize data, it is still
possible to deduce information about an individual depending on how \(x\) is distributed.
Differential privacy offers a more principled
approach by introducing randomness such that the output distribution
does not change much when one individual's data is changed. A common approach --- and the one \CRANpkg{dapper}
is primarily designed to address ---
is to take a summary statistic \(s\), and add noise to it
to produce a noisy summary statistic \(s_{dp}\).

While adding noise to confidential data is already a well-established practice in statistical disclosure control \citep{Dalenius1982}, differential privacy
provides a rigorous framework to specify where and how much
noise to add.
Most importantly, for the analyst, the specification of the differentially private noise mechanism can
be made available without compromising privacy and thus incorporated into subsequent analyses.

The \CRANpkg{dapper} package provides a flexible framework that can accommodate
the many different flavors of differential privacy. The main requirement is that
the privacy mechanism's log density or log probability mass can be evaluated.
However, for presentation,
in this section we focus on the earliest and most common
formulation of differential privacy, \(\epsilon\)-differential privacy (\(\epsilon\)-DP). The
\(\epsilon\) parameter is called the privacy loss budget. This parameter controls how
strong the privacy guarantee is. Larger values of \(\epsilon\) correspond to weaker
privacy guarantees which in turn means less noise being added.

We now describe the \(\epsilon\)-DP framework in more detail. For the noisy summary
statistic, we write \(s_{dp} \sim \eta(\cdot \mid x)\). Here, \(\eta\) denotes the
density of the privacy mechanism. The mechanism is said to be
\(\epsilon\)-differentially private \citep{Dwork2006} if for all values of \(s_{dp}\)
and all ``neighboring'' databases \((x,x') \in \mathcal{X}^n \times \mathcal{X}^n\) differing
by one record (specifically, we consider \(d(x,x') \leq 1\) where \(d\) is the Hamming distance), the probability ratio is bounded:
\begin{equation*}
\dfrac{\eta(s_{dp} \mid x)}{\eta(s_{dp} \mid x')} \leq \exp(\epsilon), \quad \epsilon > 0.
\end{equation*}

The differential privacy framework is used to create and verify privacy
mechanisms. One such mechanism is the \emph{Laplace mechanism}. Given a statistic
\(s: \mathcal{X}^n \to \mathbb{R}^m\), the Laplace mechanism constructs the
privatized statistic
\begin{equation*}
s_{dp} := s(x) + u,
\end{equation*}
where the coordinates of \(u\) are independent Laplace random variables. Define
the \(\ell_1\)-global sensitivity of \(s\) as
\begin{equation*}
\Delta_1(s) :=
\max_{\substack{(x,x') \in \mathcal{X}^n \times \mathcal{X}^n \\ d(x,x') \leq 1}}
\lVert s(x) - s(x') \rVert_1.
\end{equation*}
If each \(u_j \sim \operatorname{Lap}(0, \Delta_1(s) / \epsilon)\), then \(s_{dp}\)
is \(\epsilon\)-differentially private \citep{Dwork2006}. Example 3 applies the Laplace
mechanism to linear regression.
Other common noise-adding mechanisms include the Gaussian and the
discrete Gaussian mechanisms, which also add noise scaled to the sensitivity. \footnote{The
  Gaussian mechanisms require different and more general notions of DP than \(\epsilon\)-DP,
  in particular, zCDP \citep{Bun2016} and \((\epsilon, \delta)\)-DP \citep{Dwork2006p}.
  Example 2 will consider \((\epsilon, \delta)\)-differential privacy, which
  extends \(\epsilon\)-differential privacy to the case where the ratio bound
  can fail with probability governed by \(\delta\). More specifically, we say
  a privacy mechanism, \(\mathcal{M}\), satisfies \((\epsilon, \delta)\)-differential privacy if
  for all neighboring databases where \(d(x,x') \leq 1\), we have
  \[P(\mathcal{M}(x) \in S) \leq \exp(\epsilon) P(\mathcal{M}(x') \in S) + \delta\]
  for any \(S \subseteq \text{Range}(\mathcal{M})\) and \(\delta \in [0,1]\). Note
  setting \(\delta = 0\) gives us back the pure \(\epsilon\)-differential privacy
  condition.}

\section{Methodology}\label{sec:methodology}

Given privatized data, \(s_{dp}\), the goal of Bayesian inference is to learn the
private posterior distribution \(p(\theta \mid s_{dp})\). Since the observed likelihood,
\(p(s_{dp} \mid \theta)\), often has no simple closed-form expression \citep{NIPS2010_sherry}, most standard approaches
do not apply. To conduct privacy-aware Bayesian inference, the \CRANpkg{dapper} package implements
the data augmentation algorithm introduced in \citet{Ju2022}, which allows us to sample from \(p(\theta \mid s_{dp})\)
without knowing a closed-form expression proportional to \(p(s_{dp} \mid \theta)\).
The algorithm considers the joint distribution \(p(\theta, x \mid s_{dp})\) and
alternates between sampling from the two conditional distributions \(p(\theta \mid x, s_{dp})\)
and \(p(x \mid \theta, s_{dp})\).

Since \(s_{dp}\) is derived from \(x\), we have \(p(\theta \mid x, s_{dp}) = p(\theta \mid x)\), which
is the usual posterior distribution given the confidential data \(x\). The \CRANpkg{dapper}
package assumes that the user has access to a sampler for \(p(\theta \mid x)\). This sampler
may either draw directly from \(p(\theta \mid x)\), such as under posterior conjugacy, or
apply an MCMC transition kernel that preserves \(p(\theta \mid x)\).
For the second conditional distribution, \(p(x \mid \theta, s_{dp})\) may
only be known up to a constant. The \CRANpkg{dapper} package samples from this distribution by
running a Gibbs-like sampler: similar to a Gibbs sampler, each of the \(n\) components of \(x\) is individually
updated. However, unlike a standard Gibbs sampler, each component is updated
using a Metropolis--Hastings algorithm. This method is sometimes called a Metropolis-within-Gibbs sampler \citep{Robert2004}.

In some cases, sampling from \(p(x \mid \theta, s_{dp})\) can be made more efficient
when the privacy mechanism can be written as a function of \(s_{dp}\) and
a sum consisting of contributions from each individual record. More precisely, we say the privacy mechanism satisfies
the \emph{record-additivity} property if
\begin{equation*}
\eta(s_{dp} \mid x) = g\left(s_{dp}, t(x, s_{dp}) \right) = g\left(s_{dp}, \sum_{i=1}^{n}t_i(x_i, s_{dp}) \right),
\end{equation*}
for some known and tractable functions \(g, t_1, \ldots, t_n\) where \(t(x, s_{dp}) := \sum_{i=1}^{n} t_i(x_i, s_{dp})\).
The sample mean is an example of a summary statistic satisfying record-additivity where \(t_i(x_i, s_{dp}) = x_i\).
Record-additivity is not required for correctness. However, it allows the statistic to be updated in \(O(1)\) time after replacing a single record, rather than recomputed over all \(n\) records.

The following pseudocode shows how to generate iteration \(t+1\) from iteration \(t\)
in the data augmentation algorithm:

\begin{enumerate}
\def\labelenumi{\arabic{enumi}.}
\tightlist
\item
  Sample \(\theta^{t+1}\) from \(p(\cdot \mid x^{(t)})\).
\item
  Sample from \(p(x \mid \theta^{t+1}, s_{dp})\) using a three-step process

  \begin{itemize}
  \tightlist
  \item
    Propose \(x_{i}^{*} \sim f(\cdot \mid \theta^{t+1})\).
  \item
    Assuming \(s\) satisfies the record-additive property, perform the
    update \(t(x^*, s_{dp}) = t(x,s_{dp}) - t_i(x_i,s_{dp}) + t_{i}(x_i^*, s_{dp})\).
  \item
    Accept the proposed state with probability \(\alpha(x_i^* \mid x_i, x_{-i}, \theta^{t+1})\)
    given by:
  \end{itemize}

  \begin{equation*}
\begin{aligned}
\alpha(x_i^* \mid x_i, x_{-i}, \theta^{t+1}) &= \min \left\{ 1, \dfrac{\eta(s_{dp} \mid s(x_i^*, x_{-i}))}{\eta(s_{dp} \mid s(x_i, x_{-i}))} \right\}\\
&= \min \left\{ 1, \dfrac{g(s_{dp}, t(x^*, s_{dp}))}{g(s_{dp}, t(x,s_{dp}))} \right\}.
\end{aligned}
\end{equation*}
\end{enumerate}

Theoretical results such as bounds on the acceptance probability as well as
results on ergodicity can be found in \citet{Ju2022}.

\section{The structure of dapper}\label{sec:structure}

The \CRANpkg{dapper} package is structured around the two functions \texttt{dapper\_sample()} and
\texttt{new\_privacy()}. The function \texttt{dapper\_sample()} is used to generate MCMC draws from
the private posterior. Since constructing the private posterior requires
a large set of inputs, the process of setting up the
sampler uses a \texttt{privacy} S3 object to encapsulate
all information about the data-generating process. The role of \texttt{new\_privacy()} is to
construct \texttt{privacy} objects. This separates inputs describing the data-generating
process from inputs describing simulation parameters, which decreases
the chance for input-related bugs.

Utility functions facilitating work with count data are also included.
These center around the probability mass function and random number generators of the
discrete Gaussian and discrete Laplace distributions and are described
in more detail in Section \hyperref[sec:dp_count]{4.4}.

\subsection{Software architecture}\label{sec:architecture}

\begin{figure}[H]

{\centering \includegraphics[width=0.75\linewidth,alt={Workflow diagram connecting the four model functions to new_privacy, dapper_sample, posterior draws, and diagnostic tools.}]{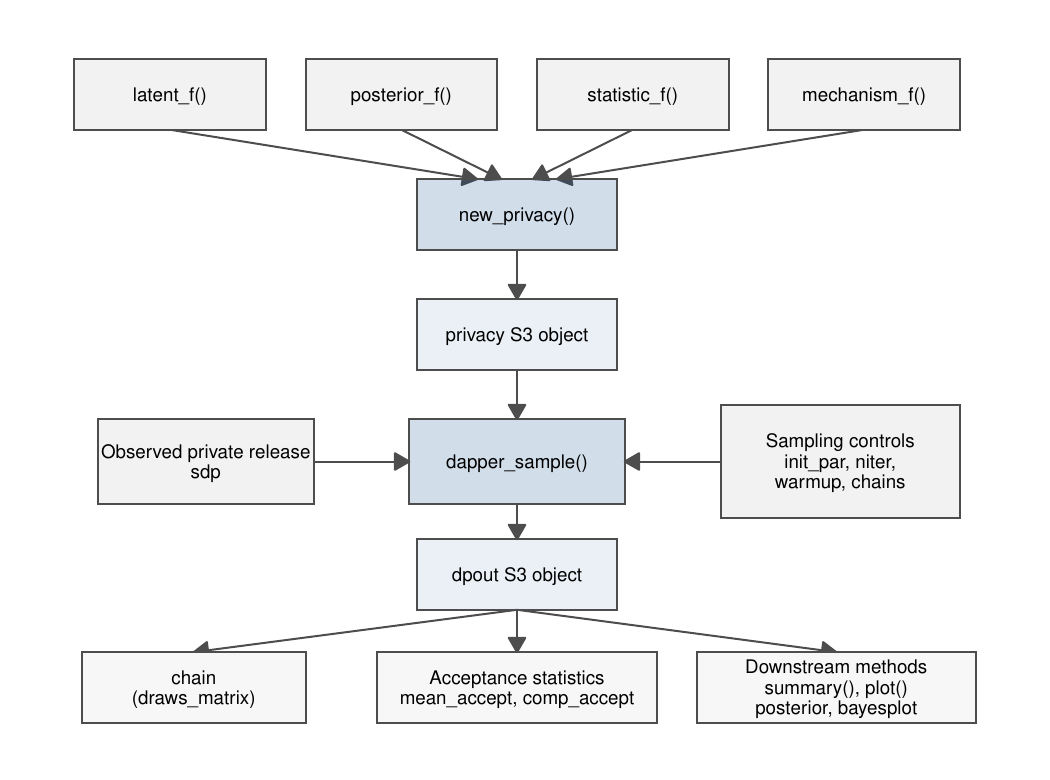} 

}

\caption{Software architecture and typical workflow of dapper. User-defined model components are assembled into a privacy object, combined with the observed private release and sampling controls by the main sampler, and returned as a dpout object for posterior analysis and diagnostics.}\label{fig:software-architecture}
\end{figure}

Figure \ref{fig:software-architecture} summarizes the typical
\CRANpkg{dapper} workflow. The four user-defined components describe the
statistical model and privacy mechanism and are validated and collected by
\texttt{new\_privacy()} in a \texttt{privacy} S3 object. This separation keeps the model
specification independent of the observed private release and the MCMC controls,
allowing a component such as \texttt{mechanism\_f()} to be replaced without rewriting
the remainder of the analysis.

The \texttt{dapper\_sample()} function combines the \texttt{privacy} object with the observed
release \texttt{sdp}, an initial parameter value, and the requested simulation settings.
It returns a \texttt{dpout} S3 object containing posterior draws and acceptance-rate
diagnostics. Its \texttt{chain} component is a \texttt{draws\_matrix}, so the result can be
summarized and visualized using the package's S3 methods or passed directly to
\CRANpkg{posterior} and \CRANpkg{bayesplot}.

\subsection{Privacy model}\label{sec:privacy_model}

Creating a privacy model is done using the \texttt{new\_privacy()} constructor. The
main arguments consist of the four components as outlined in the methodology
section.

\begin{verbatim}
new_privacy(posterior_f = NULL, latent_f = NULL, mechanism_f = NULL, 
            statistic_f = NULL, npar = NULL, varnames = NULL)
\end{verbatim}

To minimize the potential for bugs, the four main components must meet the requirements described below:

\begin{itemize}
\item
  \texttt{posterior\_f()} is a function which makes a one-step draw from the posterior given the imputed confidential data. It has
  the syntax \texttt{posterior\_f(dmat,\ theta)}. Here \texttt{dmat} is a numeric matrix representing the confidential database
  and \texttt{theta} is a numeric vector which serves as the initialization point for a one sample draw.
  The simplest case is when a conjugate prior yields a posterior from which
  \texttt{posterior\_f()} can draw independent parameter samples.
  If such a prior does not exist or is not appropriate, then \texttt{posterior\_f()} can be
  one or more steps of any Markov chain that starts with the current parameter value,
  and has this posterior as its stationary distribution.
  It is tempting to construct this by wrapping an MCMC sampler generated from other
  R packages (e.g.~\CRANpkg{rstan}, \CRANpkg{fmcmc} \citep{fmcmc_pkg}, \CRANpkg{adaptMCMC} \citep{adaptMCMC_pkg}); however, this requires caution.
  Such package samplers often use the initial burn-in samples to tune MCMC parameters,
  so that the resulting map from the input of \texttt{posterior\_f()} to its output no longer satisfies
  detailed balance (concretely, we require that if the current parameter sample comes
  from the posterior, then so does the output of \texttt{posterior\_f()}). Additionally,
  these samplers can have a large initialization overhead and implicit logic
  that makes reasoning about correctness difficult under little or no burn-in. For example,
  it is not immediately clear from the \CRANpkg{rstan} user manual how a sampler behaves
  when using no burn-in. For these reasons,
  we recommend sticking with conjugate priors when possible, as they will be quick and avoid
  serious undetected semantic errors that arise from specific implementation details of other R packages.
  If a non-conjugate prior is required, we recommend constructing \texttt{posterior\_f()}
  using a lightweight, well-documented package such as \CRANpkg{fmcmc}. When
  \texttt{posterior\_f()} uses an MCMC sampler rather than a direct conjugate draw, each
  call may perform one or more transitions targeting the conditional posterior.
  This may include an adaptive sampler, provided that its adaptation scheme is
  valid for the overall MCMC algorithm. The computational implications of choosing
  the number of inner transitions are discussed in Section \hyperref[sec:benchmarks]{6.1}.
\item
  \texttt{latent\_f()} is a function that generates data from the latent model. Its syntax
  must be \texttt{latent\_f(theta)}, where \texttt{theta} is a numeric vector representing the model
  parameters, and it must work with the \texttt{init\_par} argument of \texttt{dapper\_sample()}.
  The function returns an \(n \times p\) numeric matrix of i.i.d. draws from
  \(f(\cdot \mid \theta)\), with one proposed replacement for each record. During the
  subsequent sweep, row \(i\) is used as the proposal when updating \(x_i\). Generating
  all proposals in one call also permits vectorized implementations. The matrix
  requirement is strict, so even if \(p = 1\), \texttt{latent\_f()} should return an
  \(n \times 1\) matrix rather than a vector of length \(n\).
\item
  \texttt{mechanism\_f()} returns the log density or log probability mass of the observed
  privatized statistic \texttt{sdp}, evaluated at \texttt{sx}. Terms that are constant with respect
  to \texttt{sx} may be omitted. Its syntax is
  \texttt{mechanism\_f(sdp,\ sx)}, where \texttt{sdp} represents the observed privatized statistic
  and \texttt{sx} contains the summed record-level contributions. Both arguments must be
  numeric vectors or matrices and must appear in the stated order. The return
  value must be a scalar.
\item
  \texttt{statistic\_f()} returns the contribution of one latent record to the
  record-additive statistic. Its syntax must be \texttt{statistic\_f(xi,\ sdp,\ i)}, where \texttt{xi} contains the \(i\)th
  latent record, \texttt{sdp} is the observed privatized statistic, and \texttt{i} is the
  record index. The function must return a numeric vector or matrix. The returned
  contributions must have compatible dimensions so that they can be summed, and
  their sum must have the form expected by
  \texttt{mechanism\_f()}. In many applications, \(t_i\) depends only on \texttt{xi} and possibly
  \texttt{i}, so \texttt{sdp} is unused. The \texttt{sdp} argument supports more general record-additive
  statistics whose record-level contributions depend on the observed
  private release.
\item
  \texttt{npar} is an integer representing the dimension of \(\theta\).
\item
  \texttt{varnames} is an optional character vector containing the parameter names.
  These names are used to label the posterior draws and summary output. When
  supplied, it should contain one name for each of the \texttt{npar} parameters.
\end{itemize}

\subsection{Sampling}\label{sec:sampler}

The \texttt{dapper\_sample()} function essentially takes an existing
Bayesian model and extends it to handle privatized data. The
output of \texttt{dapper\_sample()} contains MCMC draws from
the private posterior. The function has the following syntax:

\begin{verbatim}
dapper_sample(data_model = NULL, sdp = NULL, init_par = NULL, seed = NULL,
              niter = 2000, warmup = floor(niter / 2), chains = 1)
\end{verbatim}

The parameters \texttt{data\_model}, \texttt{sdp}, and \texttt{init\_par} are required.
The \texttt{data\_model} input is a privacy object that is constructed
using \texttt{new\_privacy()} (see Section \hyperref[sec:privacy_model]{4.2}). The value
of \texttt{sdp} is equal to the observed noise-infused statistic. We require the
object class of \texttt{sdp} to be the same as the output of \texttt{statistic\_f()}.
For example, if \texttt{statistic\_f()} returns a matrix,
then \texttt{sdp} must also be a matrix. The provided starting value of the
chain (\texttt{init\_par}) must work with the \texttt{latent\_f()} component. An
error will be thrown if \texttt{latent\_f()} evaluated at \texttt{init\_par} does
not return a numeric matrix.

The \texttt{niter} argument specifies the total number of MCMC iterations run per
chain, including warmup, and \texttt{warmup} specifies how many initial iterations
are discarded. Each chain therefore contributes \texttt{niter\ -\ warmup} retained
post-warmup draws, and the returned \texttt{chain} contains
\texttt{chains\ *\ (niter\ -\ warmup)} draws in total. The other optional arguments are
\texttt{seed}, which sets the random seed, and \texttt{chains}, which specifies the number
of chains. Running a chain without warmup can be done by setting \texttt{warmup\ =\ 0}.
Running multiple chains can be done in parallel
using the \CRANpkg{furrr} \citep{furrr_pkg} package. Additionally, progress can be monitored
using the \CRANpkg{progressr} \citep{progressr_pkg} package. Adhering to the design philosophy
of the two packages, we leave the setup to the user so that they may
choose the most appropriate configuration for their system. The
contingency table demonstration given in Section 5 walks
through a typical setup of \CRANpkg{furrr} and \CRANpkg{progressr}.

The return value of \texttt{dapper\_sample()} is a list containing a \texttt{chain} component stored as
a \texttt{draws\_matrix} object with \texttt{chains\ *\ (niter\ -\ warmup)} rows. The
\texttt{mean\_accept} component is a matrix with \texttt{niter\ -\ warmup} rows and \texttt{chains}
columns. Each entry gives the fraction of latent record proposals accepted for
the corresponding iteration and chain. The \texttt{comp\_accept} component is an
\(n \times \texttt{chains}\) matrix giving the mean acceptance rate for each
latent record over the post-warmup iterations. The \texttt{draws\_matrix} object is described
in more detail in the \CRANpkg{posterior} \citep{posterior_pkg} package. The advantages of working
with a \texttt{draws\_matrix} object are that it can be manipulated using \CRANpkg{posterior} and is
compatible with many packages in the \CRANpkg{rstan} ecosystem. For example, a
\texttt{draws\_matrix} object can be passed directly to the popular \CRANpkg{bayesplot}
\citep{bayesplot_pkg} package. Example 1 demonstrates both interfaces explicitly. Additionally,
\CRANpkg{dapper}'s basic summary function provides the same posterior summary statistics
as those found when using \CRANpkg{rstan}. Overall, this should make working with \CRANpkg{dapper} easier
for anyone already familiar with the \CRANpkg{rstan} ecosystem.

\subsection{Privacy mechanisms for count data}\label{sec:dp_count}

The \CRANpkg{dapper} package provides several utility functions
for analyzing privatized count data. Privatized count data are common in official and public statistics.
As an example, the most prominent deployment of differential privacy for public data dissemination is the 2020 U.S. Decennial Census, consisting of a collection of tabular data products that follow a geographic hierarchy.

Pure \(\epsilon\)-differential privacy places a stringent requirement on the privacy noise that must be introduced,
which can lead to poor data quality. For this reason, alternative choices of privacy noise can be desirable.
For example, the U.S. Census Bureau adopts zero-concentrated differential privacy (zCDP), a more general criterion that permits a broader range of noise mechanisms.
The \CRANpkg{dapper} package includes probability mass and sampling functions
for the discrete Gaussian and discrete Laplace distributions \citep{Canonne2022}, both of which are commonly used in the zCDP framework.
The 2020 U.S. Decennial Census data products are protected with two sets of mechanisms that satisfy zCDP: the TopDown mechanism for the redistricting and Demographic and Housing Characteristics (DHC) files \citep{Abowd2022}, and the SafeTab mechanism for the detailed DHC files \citep{TumultLabs2022}.

Equations \eqref{eq:dgauss} and \eqref{eq:dlaplace} in the panel below give the
probability mass functions for the discrete Gaussian and discrete Laplace distributions, respectively:
\begin{equation}
P[X = x] = \dfrac{e^{-(x - \mu)^2/(2\sigma^2)}}{\sum_{y \in \mathbb{Z}} e^{-(y-\mu)^2/(2\sigma^2)}},
\label{eq:dgauss}
\end{equation}

\begin{equation}
P[X = x] = \dfrac{e^{1/t} - 1}{e^{1/t} + 1} e^{-|x|/t}.
\label{eq:dlaplace}
\end{equation}

The support of both distributions is the set of all integers. The discrete Gaussian has
two parameters \((\mu,\sigma) \in \mathbb{R} \times \mathbb{R}^+\), which govern the location and
scale, respectively. On the other hand, the discrete Laplace only has the scale parameter \(t \in \mathbb{R}^+\). The functions \texttt{ddnorm} and \texttt{rdnorm} provide the
density and sampling features for the discrete Gaussian distribution. The normalizing constant used by \texttt{ddnorm()} is cached for reuse across calls with
the same \texttt{sigma}. When \texttt{log\ =\ TRUE}, the function returns unnormalized log
probabilities. Finally, the functions
\texttt{ddlaplace} and \texttt{rdlaplace} provide similar features for the discrete Laplace distribution.
The functions \texttt{ddnorm()} and \texttt{rdnorm()} require an integer \texttt{mu}. For the discrete
Laplace distribution, \texttt{ddlaplace()} accepts any positive \texttt{scale}, while
\texttt{rdlaplace()} requires a positive integer \texttt{scale}. Due to potential floating-point attacks \citep{Mironov2012}, the implementation of these functions is not the safest and is merely meant to provide
users with a way to quickly explore the behavior of these DP mechanisms. The implementation
logic for both the discrete Gaussian and discrete Laplace mechanisms closely follows the algorithms as outlined in \citep{Canonne2022}.

\section{Examples}\label{sec:examples}

\subsection{Example 1: 2x2 contingency table (randomized response)}\label{example-1-2x2-contingency-table-randomized-response}

As a demonstration, we analyze a subset of the UC Berkeley admissions data, which is often
used as an illustrative example of Simpson's paradox \citep{Bickel1975}. The question posed is whether
the data suggest there is bias against females during the college admissions
process. We work with a \(\sim 9\%\) subsample consisting of \(N = 400\) applicants for
computational convenience. The paired tables below show admissions outcomes aggregated across six departments by sex
for this subsample. The left subtable shows the confidential data, and the right subtable shows the resulting counts after applying the randomized response privacy mechanism.

\begin{table}
\centering\caption{\label{tab:admissions-rr}The left-hand table shows the confidential admissions data, and the right-hand table shows the perturbed counts after application of the randomized response mechanism. The mechanism substantially reduces the observed difference in admission rates between female and male applicants.}

\centering
\begin{tabular}[t]{lrr}
\toprule
  & Admitted & Rejected\\
\midrule
Female & 46 & 118\\
Male & 109 & 127\\
\bottomrule
\end{tabular}
\centering
\begin{tabular}[t]{lrr}
\toprule
  & Admitted & Rejected\\
\midrule
Female & 74 & 102\\
Male & 104 & 120\\
\bottomrule
\end{tabular}
\end{table}

To see how the privacy mechanism works, we represent the record-level data as an
\(N \times 2\) matrix, with the first column indicating sex and the second
indicating admission status. Each row therefore represents one individual. We
apply randomized response independently to each binary attribute: the original
value is preserved with probability \(3/4\) and flipped with probability \(1/4\).\footnote{The randomized response scheme predates the development of differential privacy and was first described by \citet{Warner1965} as a means to reduce survey bias involving sensitive questions.}
This rule can, for example, be implemented by flipping two fair coins for each
attribute and changing the value only for one designated combination of
outcomes. Thus, for an individual recorded as male and rejected, the sex and
admission-status values are randomized separately. Applying the mechanism to
both attributes gives a privacy loss budget of at most \(\epsilon = 2\log(3)\).

To set up \CRANpkg{dapper} to analyze the anonymized admissions data, we encode
the randomized record-level data as a length-800 vector. This vector is formed
by stacking the sex column of the binary data matrix above the admission-status
column, where male and admitted take the value 1.

\begin{enumerate}
\def\labelenumi{\arabic{enumi}.}
\item
  \texttt{latent\_f}: For each individual there are four possible
  sex/status responses which can be modeled using a multinomial distribution.
  To implement draws from the multinomial distribution we use the \texttt{sample} function
  to take samples from a list containing the four possible binary vectors. Note
  the final line results in a \(400 \times 2\) matrix.

\begin{verbatim}
latent_f <- function(theta) {
  tl <- list(c(1,1), c(1,0), c(0,1), c(0,0))
  rs <- sample(tl, 400, replace = TRUE, prob = theta)
  do.call(rbind, rs)
}
\end{verbatim}
\item
  \texttt{posterior\_f}: Given the confidential data, we can derive the posterior analytically
  using a Dirichlet prior. In this example, we use a flat prior which
  corresponds to a Dirichlet(1) distribution. The code below generates samples from the Dirichlet distribution
  using random draws from the gamma distribution.

\begin{verbatim}
 posterior_f <- function(dmat, theta) {
  sex <- dmat[,1]
  status <- dmat[,2]

  #Male & Admit
  x1 <- sum(sex & status)
  x2 <- sum(sex & !status)
  x3 <- sum(!sex & status)
  x4 <- sum(!sex & !status)

  x <- c(x1, x2, x3, x4)

  t1 <- rgamma(4, x + 1, 1)
  t1/sum(t1)
}
\end{verbatim}
\item
  \texttt{statistic\_f}: Because randomized response is applied independently to the
  two attributes, the mechanism likelihood depends on how many privatized
  attributes match the corresponding latent attributes. For record \(i\), define
  \begin{equation*}
t_i(x_i, s_{dp}) =
\mathbb{1}(x_{i1} = s_{dp,i}) +
\mathbb{1}(x_{i2} = s_{dp,n+i}).
\end{equation*}
  Summing these contributions gives the total number of matching entries across
  the database.

\begin{verbatim}
statistic_f <- function(xi, sdp, i) {
  n <- length(sdp) %/% 2
  sum(xi == sdp[c(i, n + i)])
}
\end{verbatim}
\item
  \texttt{mechanism\_f}: Each privatized attribute matches its confidential value with
  probability \(3/4\) and differs with probability \(1/4\). If \texttt{sx} is the total
  number of matching entries, the log likelihood is
  \begin{equation*}
\texttt{sx}\log(3/4) + (2n - \texttt{sx})\log(1/4).
\end{equation*}

\begin{verbatim}
mechanism_f <- function(sdp, sx) {
  sx * log(3/4) + (length(sdp) - sx) * log(1/4)
}
\end{verbatim}
\end{enumerate}

Below, we load the data and create the noisy admissions table.

\begin{verbatim}
#Original UCBAdmissions data.
cnf_df <- tibble(sex = c(1, 1, 0, 0),
                 status = c(1, 0, 1, 0),
                 n = c(1198, 1493, 557, 1278)) %>% uncount(n)

set.seed(1) 
ix <- sample(1:nrow(cnf_df), 400, replace = FALSE)
cnf_df <- cnf_df[ix,]

#Answers to be randomized
ri <- as.logical(rbinom(800, 1, 1/2)) 

#Randomized answers
ra <- rbinom(sum(ri), 1, 1/2)

#Create sdp
sdp <- c(as.matrix(cnf_df))
sdp[ri] <- ra
\end{verbatim}

Once we have defined all components of the model, we can
create a new privacy model object using the \texttt{new\_privacy()} function and
feed this into the \texttt{dapper\_sample()} function. Below, we run four chains in
parallel for 6,000 total MCMC iterations per chain, discarding the first 1,000
iterations as warmup. This leaves 5,000 retained post-warmup draws per chain
and 20,000 pooled draws across the four chains.\footnote{The number of retained post-warmup draws
  used in each example is chosen for computational convenience, ensuring a reasonable effective
  sample size within a short runtime. In a formal analysis, however, a larger effective sample size is
  typically required, particularly when estimating posterior quantiles near the tails.}

\begin{verbatim}
library(dapper)
library(furrr)
plan(multisession, workers = 4)

dmod <- new_privacy(posterior_f = posterior_f,
                    latent_f    = latent_f,
                    mechanism_f = mechanism_f,
                    statistic_f = statistic_f,
                    npar     = 4,
                    varnames = c("pi_11", "pi_10", "pi_01", "pi_00"))
                  
dp_out <- dapper_sample(dmod,
                  sdp = sdp,
                  seed = 123,
                  niter = 6000,
                  warmup = 1000,
                  chains = 4,
                  init_par = rep(.25,4))
\end{verbatim}

If the runtime of \texttt{dapper\_sample()} is exceptionally long, one can
use the \texttt{progressr} package to monitor progress. The \texttt{progressor} framework
allows for a unified handling of progress bars in both the sequential and
parallel computing case.

\begin{verbatim}
library(progressr)
handlers(global = TRUE)

handlers("cli")
dp_out <- dapper_sample(dmod,
                  sdp = sdp,
                  seed = 123,
                  niter = 6000,
                  warmup = 1000,
                  chains = 4,
                  init_par = rep(.25,4))
\end{verbatim}

Results can be quickly summarized using \texttt{summary()}, which delegates to
\texttt{summarise\_draws()} from the \CRANpkg{posterior} package. The \texttt{rhat} values in the table below are close
to 1, which is consistent with convergence. They should be considered together
with the trace plots and effective sample sizes.

\begin{verbatim}
#> # A tibble: 4 x 10
#>   variable  mean median     sd    mad     q5   q95  rhat ess_bulk ess_tail
#>   <chr>    <dbl>  <dbl>  <dbl>  <dbl>  <dbl> <dbl> <dbl>    <dbl>    <dbl>
#> 1 pi_11    0.281  0.281 0.0610 0.0625 0.182  0.382  1.02     362.     818.
#> 2 pi_10    0.336  0.335 0.0638 0.0640 0.235  0.444  1.01     431.    1191.
#> 3 pi_01    0.111  0.108 0.0548 0.0563 0.0250 0.206  1.02     282.     504.
#> 4 pi_00    0.272  0.272 0.0601 0.0616 0.172  0.372  1.02     389.     875.
\end{verbatim}

\begin{figure}[H]

{\centering \includegraphics[alt={Four colored Markov chain traces for each of the four cell-probability parameters in Example 1.}]{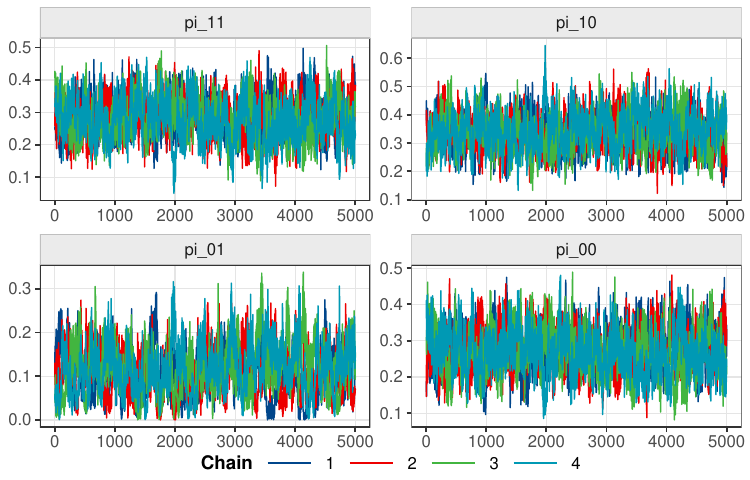} 

}

\caption{(Example 1) Trace plots created by passing the dapper draws directly to bayesplot. Each chain is represented by a distinct qualitative color. By returning posterior draws in a standard format, dapper makes it easy to use existing Bayesian diagnostic tools such as bayesplot.}\label{fig:trace-plot}
\end{figure}

Because \texttt{dp\_out\$chain} is a \texttt{draws\_matrix}, it can be passed directly to
\CRANpkg{bayesplot}. The call underlying \texttt{plot(dp\_out)} is shown explicitly in
Figure \ref{fig:trace-plot}. It is especially important to check for good mixing
with \CRANpkg{dapper} since sticky chains are likely to be produced when the
amount of injected noise is high. See Section \hyperref[sec:mixing]{6.2} for a more detailed
explanation.

\begin{verbatim}
bayesplot::mcmc_trace(dp_out$chain) +
  ggsci::scale_color_lancet() +
  paper_theme +
  guides(color = guide_legend(title = "Chain", title.position = "left", nrow = 1))
\end{verbatim}

To see if there is evidence of gender bias, we can look at the odds ratio.
Specifically, we compare the odds of a male applicant being admitted with
those of a female applicant. A higher odds ratio would indicate a bias favoring males. The
\texttt{mutate\_variables()} and \texttt{subset\_draws()} functions from \CRANpkg{posterior} can
be used to create and select this derived posterior quantity without converting
the \CRANpkg{dapper} output to another format.

\begin{verbatim}
odds_ratio_draws <- posterior::mutate_variables(
  dp_out$chain,
  odds_ratio = (pi_11 * pi_00) / (pi_10 * pi_01)
) |>
  posterior::subset_draws(variable = "odds_ratio")

posterior::summarise_draws(odds_ratio_draws)
\end{verbatim}

\begin{verbatim}
#> # A tibble: 1 x 10
#>   variable    mean median    sd   mad    q5   q95  rhat ess_bulk ess_tail
#>   <chr>      <dbl>  <dbl> <dbl> <dbl> <dbl> <dbl> <dbl>    <dbl>    <dbl>
#> 1 odds_ratio  6.83   2.12  82.0  1.89 0.472  16.7  1.03     245.     568.
\end{verbatim}

\begin{figure}[H]

{\centering \includegraphics[alt={Red density estimate of the randomized-response odds ratio, with a central 80 percent interval and a vertical reference line at one.}]{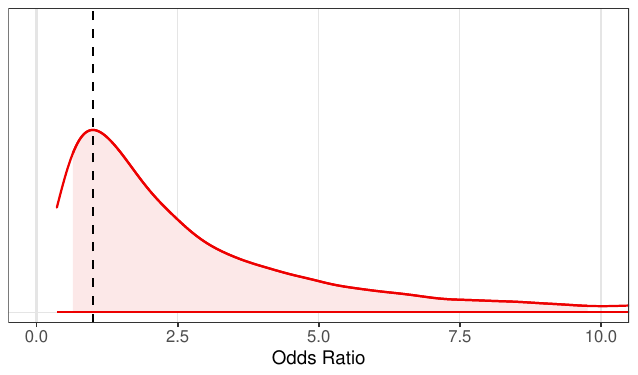} 

}

\caption{(Example 1) Posterior density estimate for the odds ratio, displayed over the range 0 to 10 and based on 20,000 retained post-warmup draws (5,000 from each of four chains). The light-red region denotes the central 80\% interval, and the dashed vertical line marks the null value of 1. The posterior is strongly right-skewed, and its central 80\% interval includes the null value of 1.}\label{fig:post-or-density}
\end{figure}

Figure \ref{fig:post-or-density} visualizes the derived odds ratio using
\CRANpkg{bayesplot}. The large odds ratio values would seem to indicate that there is
bias favoring males. Stratifying the data by university department reveals
Simpson's paradox. The within-department comparisons reverse the aggregate
pattern and tend to favor female applicants.

\begin{verbatim}
previous_color_scheme <- attr(bayesplot::color_scheme_get(), "scheme_name")
bayesplot::color_scheme_set(privatized_color_scheme)

odds_ratio_plot <- bayesplot::mcmc_areas(
  odds_ratio_draws,
  prob = 0.8,
  prob_outer = 0.95,
  point_est = "none"
) +
  geom_vline(xintercept = 1, linetype = "dashed", linewidth = 0.5) +
  coord_cartesian(xlim = c(0, 10)) +
  labs(x = "Odds Ratio", y = NULL) +
  paper_theme +
  theme(
    axis.text.y = element_blank(),
    axis.ticks.y = element_blank()
  )

bayesplot::color_scheme_set(previous_color_scheme)
odds_ratio_plot
\end{verbatim}

\begin{figure}[H]

{\centering \includegraphics[alt={Two panels compare confidential-data odds-ratio densities in blue with privacy-aware and naive randomized-response analyses in solid and dashed red.}]{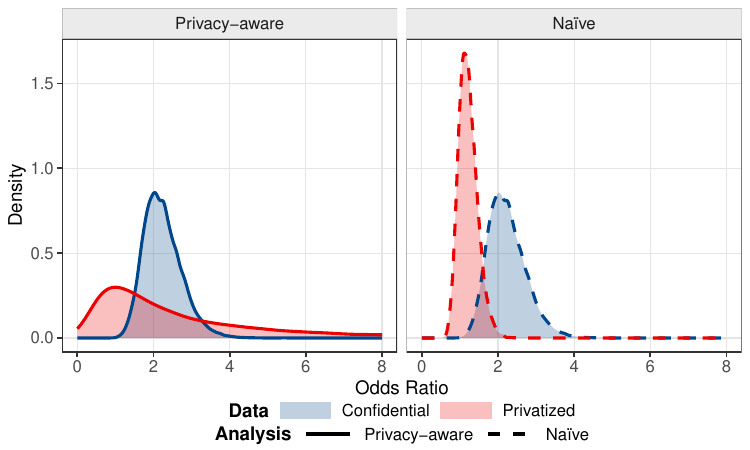} 

}

\caption{(Example 1) Posterior distributions of the odds ratio
(admission of males vs. females). Blue denotes confidential data and red denotes
privatized data. Solid lines denote privacy-aware inference using dapper; dashed
lines denote naïve inference that treats the privatized data as noise-free.
Accounting for randomized response substantially increases posterior uncertainty relative to the naïve analysis.}\label{fig:post-or-compare}
\end{figure}

For comparison, we run a standard Bayesian analysis on the
noise-infused table ignoring the privacy mechanism. This will
correspond exactly to the model defined in the \texttt{posterior\_f} component.
Figure \ref{fig:post-or-compare} shows a density estimate for the odds ratio
under the confidential and noisy data. The posterior
distribution for the odds ratio under the noisy data
is shifted significantly, indicating a large degree of bias.
The left-hand plot in Figure \ref{fig:post-or-compare} shows that the MAP estimate from \CRANpkg{dapper}
is similar to that in the case of the confidential data.
The width of the posterior is also much larger since
it properly accounts for the uncertainty due to the privacy mechanism. This
illustrates the dangers of ignoring the privacy mechanism: a naïve
analysis is not only biased but also severely underestimates the
uncertainty associated with the odds ratio estimate.

\subsection{Example 2: 2x2 contingency table (discrete Gaussian)}\label{example-2-2x2-contingency-table-discrete-gaussian}

To highlight the flexibility of \CRANpkg{dapper}, we reanalyze Example 1
with a different privacy mechanism. Here we take inspiration from the 2020 U.S. Decennial Census,
which deployed a novel privacy protection system for count data.
In particular, the discrete Gaussian distribution is used in SafeTab.
In our admissions data example, this privacy mechanism works by injecting
noise into the total cell counts given in the 2x2 table. The
randomized response scheme, in contrast, injects noise at the record level.
The \CRANpkg{dapper} package accommodates both the randomized response and
the discrete Gaussian mechanisms, allowing us to compare the impact
of the two approaches.

These mechanisms also illustrate two different trust models. Randomized
response is a local differential privacy mechanism in which each individual's
attributes are privatized before being made available to the data curator. On
the other hand, the discrete Gaussian mechanism operates under central
differential privacy, in which a trusted curator collects the confidential
records and adds noise to the aggregate counts before release. Local
differential privacy therefore protects the data from the curator as well as
downstream users, but this stronger protection can entail substantially greater
statistical cost for many estimation problems \citep{Duchi2018, BarberDuchi2014}.

To begin comparing the two approaches, we need to set the privacy parameters.
Since the two frameworks use different metrics, there does not exist
a direct comparison. However, there is a direct relationship between
zCDP and \((\epsilon, \delta)\)-differential privacy. The latter framework
is a relaxed version of \(\epsilon\)-DP where the ratio bound only holds
in probability. For our comparison, we will set \(\delta = 10^{-10}\) which
is the value used in the 2020 Decennial Census.
For this value of \(\delta\), setting the scale parameter in the discrete Gaussian
distribution to \(\sigma = 6.32\) will guarantee \((2\log(3), 10^{-10})\)-differential privacy.\footnote{\(\dfrac{1}{2} \epsilon^2\)-concentrated differential privacy implies \(\left(\dfrac{1}{2}\epsilon^2 + \epsilon \sqrt{2\log(1/\delta)}, \delta\right)\)-differential privacy \citep{Bun2016}.}

\begin{table}
\centering\caption{\label{tab:admissions-dg}The left-hand table shows the confidential admissions data, and the right-hand table shows the perturbed data resulting from the discrete Gaussian mechanism with the scale parameter set to 6.32. The perturbed counts remain close to the confidential counts and retain the higher observed admission rate for male applicants.}

\centering
\begin{tabular}[t]{lrr}
\toprule
  & Admitted & Rejected\\
\midrule
Female & 46 & 118\\
Male & 109 & 127\\
\bottomrule
\end{tabular}
\centering
\begin{tabular}[t]{lrr}
\toprule
  & Admitted & Rejected\\
\midrule
Female & 47 & 110\\
Male & 110 & 131\\
\bottomrule
\end{tabular}
\end{table}

For the public, the 2020 Decennial Census only shows aggregate cell counts along with
the true total count of those cells. In our example,
this means the Census would have released the right-hand
table along with the fact that \(N = 400\) in the original
table. Thus, it is natural to let \(s_{dp}\) be the vector of cell counts. As in
Example 1, we imagine the latent database as a
\(400 \times 2\) binary matrix. Below,
we describe the process for analyzing the privatized
data using \CRANpkg{dapper}. Since the latent process
and posterior are the same as Example 1, we only describe
how to construct \texttt{statistic\_f} and \texttt{mechanism\_f}.

\begin{enumerate}
\def\labelenumi{\arabic{enumi}.}
\item
  \texttt{statistic\_f}: The private summary statistic \(s_{dp}\) can be written as a record-additive
  statistic using the indicator vectors \((1,0,0,0), (0,1,0,0), (0,0,1,0)\) and \((0,0,0,1)\).
  These four vectors correspond to the four possible cells.

\begin{verbatim}
statistic_f <- function(xi, sdp, i) {
  if(xi[1] & xi[2]) {
    c(1,0,0,0)
  } else if (xi[1] & !xi[2]) {
    c(0,1,0,0)
  } else if (!xi[1] & xi[2]) {
    c(0,0,1,0)
  } else {
    c(0,0,0,1)
  }
}
\end{verbatim}
\item
  \texttt{mechanism\_f}: The privacy mechanism is a discrete Gaussian distribution centered
  at 0.

\begin{verbatim}
mechanism_f <- function(sdp, sx) {
  sum(dapper::ddnorm(sdp - sx, mu = 0, sigma = 6.32, log = TRUE))
}
\end{verbatim}
\end{enumerate}

The summary table below shows the results from one chain run for 2,000 total
MCMC iterations. The first 1,000 iterations are discarded as warmup, leaving
1,000 retained post-warmup draws.

\begin{verbatim}
summary(dp_out)
\end{verbatim}

\begin{verbatim}
#> # A tibble: 4 x 10
#>   variable  mean median     sd    mad     q5   q95  rhat ess_bulk ess_tail
#>   <chr>    <dbl>  <dbl>  <dbl>  <dbl>  <dbl> <dbl> <dbl>    <dbl>    <dbl>
#> 1 pi_11    0.278  0.277 0.0247 0.0234 0.238  0.321 1.00      624.     778.
#> 2 pi_10    0.326  0.325 0.0269 0.0263 0.282  0.372 1.00      586.     546.
#> 3 pi_01    0.120  0.120 0.0213 0.0214 0.0855 0.156 1.000     450.     670.
#> 4 pi_00    0.276  0.276 0.0264 0.0266 0.234  0.321 1.00      644.     680.
\end{verbatim}

\begin{figure}[H]

{\centering \includegraphics[alt={Red odds-ratio density estimates compare randomized response with discrete Gaussian privacy noise using dashed and solid lines.}]{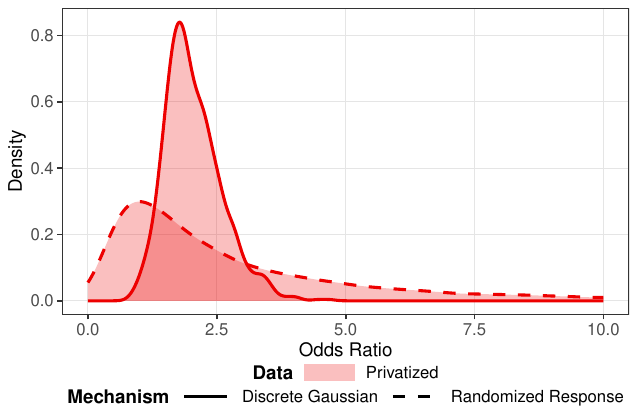} 

}

\caption{(Example 2) Privacy-aware posterior density estimates for the odds ratio based on privatized data (red). Dashed and solid lines denote the randomized response and discrete Gaussian mechanisms, respectively. The estimates are based on 20,000 retained post-warmup draws for randomized response (5,000 from each of four chains) and 1,000 retained post-warmup draws from one chain for the discrete Gaussian mechanism. For these releases, the discrete Gaussian mechanism yields a more concentrated odds-ratio posterior than randomized response.}\label{fig:post-or-density-dg}
\end{figure}

Figure \ref{fig:post-or-density-dg} juxtaposes the private posterior under the
randomized response (dashed line) and discrete Gaussian (solid line) mechanisms. The
comparison suggests that using discrete Gaussian noise leads to slightly less posterior
uncertainty and a mode closer to that of the confidential-data posterior.

\begin{figure}[H]

{\centering \includegraphics[alt={Two panels compare confidential-data odds-ratio densities in blue with privacy-aware and naive discrete-Gaussian analyses in solid and dashed red.}]{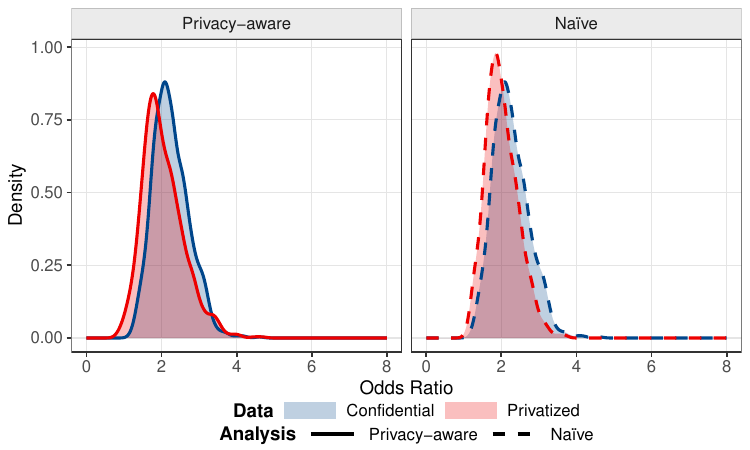} 

}

\caption{(Example 2) Posterior distributions of the odds ratio
(admission of males vs. females). Blue denotes confidential data and red denotes
privatized data. Solid lines denote privacy-aware inference using dapper; dashed
lines denote naïve inference that treats the privatized data as noise-free.
The privacy-aware posterior closely resembles the posterior obtained from the confidential data.}\label{fig:post-or-compare-dg}
\end{figure}

Additionally, comparing the plots in Figures
\ref{fig:post-or-compare} and \ref{fig:post-or-compare-dg} shows that the
discrete Gaussian mechanism induces considerably less bias under the naïve analysis.

\subsection{Example 3: linear regression}\label{example-3-linear-regression}

In this section we apply \CRANpkg{dapper} to reconstruct an
example presented in \citet{Ju2022}. In it, they apply a Laplace privacy
mechanism to a sufficient summary statistic for a linear regression model.
Let \(\{(x_i,y_i)\}_{i=1}^{n}\) be the original, confidential data with \(x_i \in \mathbb{R}^2\).
They assume the true data-generating process follows the model

\begin{equation*}
\begin{aligned}
y &= -1.79 -2.89x_1 -0.66x_2 + \epsilon\\
\epsilon &\sim N(0,\sigma^2), \quad \sigma^2 = 2\\
\binom{x_1}{x_2} &\sim N_{2}(\mu, I_2)\\
\mu &= \binom{0.9}{-1.17}.
\end{aligned}
\end{equation*}

In most settings involving linear regression, the covariates are assumed to be
fixed, known constants. Thus, the formulation above is a departure from the norm since
we are assuming a random design matrix. More details on why this framing is necessary
will be provided later when describing the latent model. The paper considers the scenario where one desires to publicly release the
sufficient summary statistics

\begin{equation*}
s(x,y) = (x^Ty, y^Ty, x^Tx).
\end{equation*}

This summary statistic satisfies the additive record property since \(s(x,y) = \sum_{i=1}^{n} t(x_i, y_i)\),
where
\begin{equation*}
t(x_i,y_i) = ((x_{i})^T y_i, y_i^2, (x_{i})^T x_i).
\end{equation*}
To make the statistic compliant with
the \(\epsilon\)-DP criterion, it is
necessary to bound the value of the statistic
(i.e.~the statistic must have finite global sensitivity). This is accomplished
by clamping the data, which prevents any data point from being too ``unique.''
More precisely, we define
the clamping function \([z] := \min\{\max\{z,-10\}, 10\}\) which truncates a value
\(z\) so that it falls into the interval \([-10,10]\). Furthermore, we let \(\tilde{z} := [z]/10\)
denote the normalized clamped value of \(z\). The clamped statistic is

\begin{equation*}
t(x_i,y_i) = ((\tilde{x}^{i})^T \tilde{y}_i, \tilde{y}_i^2, (\tilde{x}_{i})^T \tilde{x}_i).
\end{equation*}

Ignoring duplicate entries in the symmetric matrix \(x^Tx\), the statistic has \(\ell_1\)-sensitivity \(\Delta = p^2 + 4p + 3\),
where \(p\) is the number of predictors in the regression model (in this example \(p = 2\))
.\footnote{The original paper, \citet{Ju2022}, contains a computation error and mistakenly uses \(\Delta = p^2 + 3p + 3\).}
Using the Laplace mechanism, \(\epsilon\)-DP can thus be achieved by adding i.i.d. Laplace\((0, \Delta/\epsilon)\)
error to each unique entry. For techniques that yield a tighter sensitivity bound,
see \citet{Awan2021}.

\begin{enumerate}
\def\labelenumi{\arabic{enumi}.}
\item
  \texttt{latent\_f}: Because the summary statistic encodes information about the design
  matrix and the privacy mechanism adds noise to that statistic, we cannot treat the
  design matrix as fixed and known. Hence, to draw a sample from the latent data-generating
  process we use the relation \(f(x,y) = f(x)f(y \mid x)\). In this formulation,
  it is necessary to specify a distribution on the covariates \(x\).

\begin{verbatim}
latent_f <- function(theta) {
  xmat <- MASS::mvrnorm(50 , mu = c(.9,-1.17), Sigma = diag(2))
  y <- cbind(1,xmat) %*% theta + rnorm(50, sd = sqrt(2))
  cbind(y,xmat)
}
\end{verbatim}
\item
  \texttt{posterior\_f}: Given confidential data \(X\), we can derive the posterior analytically
  using a normal prior on \(\beta\).
  \begin{equation*}
\begin{aligned}
\beta &\sim N_{p+1}(0, \tau^2 I_{p+1})\\
\beta \mid x,y &\sim N(\mu_n, \Sigma_n)\\
\Sigma_n &= (x^Tx/\sigma^2 + I_{p+1}/\tau^2)^{-1}\\
\mu_n &= \Sigma_n(x^Ty)/\sigma^2
\end{aligned}
\end{equation*}
  In the example, we use \(\sigma^2 = 2\) and \(\tau^2 = 4\).

\begin{verbatim}
posterior_f <- function(dmat, theta) {
  x <- cbind(1,dmat[,-1])
  y <- dmat[,1]

  ps_s2 <- solve((1/2) * t(x) %*% x + (1/4) * diag(3))
  ps_m <- ps_s2 %*% (t(x) %*% y) * (1/2)

  MASS::mvrnorm(1, mu = ps_m, Sigma = ps_s2)
}
\end{verbatim}
\item
  \texttt{statistic\_f}: The summary statistic contains duplicate
  entries. We can considerably reduce the dimension of the
  statistic by only considering unique entries. The \texttt{clamp\_data}
  function is used to bound the statistic to give a finite
  global sensitivity.

\begin{verbatim}
clamp_data <- function(dmat) {
  pmin(pmax(dmat,-10),10) / 10
}

statistic_f <- function(xi, sdp, i) {
  xic <- clamp_data(xi)
  ydp <- xic[1]
  xdp <- cbind(1,t(xic[-1]))

  s1 <- t(xdp) %*% ydp
  s2 <- t(ydp) %*% ydp
  s3 <- t(xdp) %*% xdp

  ur_s1 <- c(s1)
  ur_s2 <- c(s2)
  ur_s3 <- s3[upper.tri(s3,diag = TRUE)][-1]
  c(ur_s1,ur_s2,ur_s3)
}
\end{verbatim}
\item
  \texttt{mechanism\_f}: The privacy mechanism
  adds Laplace\((0, \Delta/\epsilon)\) error to each unique entry
  of the statistic. In this example, \(\Delta = 15\) and \(\epsilon = 10\).

\begin{verbatim}
mechanism_f <- function(sdp, sx) {
  sum(VGAM::dlaplace(sdp - sx, 0, 15/10, log = TRUE))
}
\end{verbatim}
\end{enumerate}

First, we simulate fake data using the aforementioned privacy mechanism.
In the example, we use \(n = 50\) observations.

\begin{verbatim}
deltaa <- 15
epsilon <- 10
n <- 50

set.seed(1)
xmat <- MASS::mvrnorm(n, mu = c(.9,-1.17), Sigma = diag(2))
beta <- c(-1.79, -2.89, -0.66)
y <- cbind(1,xmat) %*% beta + rnorm(n, sd = sqrt(2))

#clamp the confidential data in xmat
dmat <- cbind(y,xmat)
sdp <- apply(
  sapply(seq_len(nrow(dmat)), 
  function(i) {
    statistic_f(dmat[i, ], NULL, i)
  }),
  1,
  sum
)

#add Laplace noise 
sdp <- sdp + VGAM::rlaplace(length(sdp), location = 0, scale = deltaa/epsilon)
\end{verbatim}

We construct a privacy model using the \texttt{new\_privacy()} function and run one chain
for 25,000 total MCMC iterations. The first 1,000 iterations are discarded as
warmup, leaving 24,000 retained post-warmup draws.

\begin{verbatim}
library(dapper)
set.seed(123)

dmod <- new_privacy(posterior_f = posterior_f,
                    latent_f    = latent_f,
                    mechanism_f = mechanism_f,
                    statistic_f = statistic_f,
                    npar     = 3,
                    varnames = c("beta0", "beta1", "beta2"))



dp_out <- dapper_sample(dmod,
                        sdp = sdp,
                        niter = 25000,
                        warmup = 1000,
                        chains = 1,
                        init_par = rep(0,3))
\end{verbatim}

The output of the MCMC run is reported below.

\begin{verbatim}
summary(dp_out)
\end{verbatim}

\begin{verbatim}
#> # A tibble: 3 x 10
#>   variable   mean median    sd   mad    q5   q95  rhat ess_bulk ess_tail
#>   <chr>     <dbl>  <dbl> <dbl> <dbl> <dbl> <dbl> <dbl>    <dbl>    <dbl>
#> 1 beta0    -0.939 -0.852  1.52  1.52 -3.59  1.43  1.01     439.     869.
#> 2 beta1    -1.87  -2.21   1.50  1.20 -3.73  1.23  1.00     133.     186.
#> 3 beta2     0.760  0.708  1.37  1.51 -1.39  3.02  1.00     165.     608.
\end{verbatim}

For comparison, we consider a Bayesian analysis in which the design matrix
is a fixed, known constant and \(\sigma^2\) is known. This fixed-design analysis
does not exactly correspond to the model used in \texttt{posterior\_f()}, but it serves
as the naïve analysis. However, under a prior in which the regression and
design-matrix parameters are independent, the design-matrix parameters can be
profiled out. As such, inference
for \(\beta\) can be carried out by conditioning on the observed covariates.
Using the diffuse prior \(f(\beta) \propto 1\) leads to a normal posterior.
\begin{equation*}
\begin{aligned}
\beta \mid x,y, \sigma^2 &\sim N(\hat{\beta}, \hat{\Sigma})\\
\hat{\beta} &= (x^Tx)^{-1}x^Ty\\
\hat{\Sigma} &= \sigma^{2}(x^Tx)^{-1}.
\end{aligned}
\end{equation*}

The posterior can be written as a function of \(s(x,y)\), but since
we only have access to the noisy version \(s_{dp}\) we
attempt to reconstruct it by extracting the
relevant entries, as shown below.

\begin{verbatim}
#x^Ty
s1 <- sdp[1:3]

#y^Ty
s2 <- sdp[4]

#x^Tx
s3 <- matrix(0, nrow = 3, ncol = 3)
s3[upper.tri(s3, diag = TRUE)] <- c(n, sdp[5:9])
s3[lower.tri(s3)] <- s3[upper.tri(s3)]
\end{verbatim}

Because of the injected privacy noise, the reconstructed
\((x^Tx)^{-1}\) matrix is not positive definite. As a naïve solution, we use
\texttt{nearest\_spd()} from the \CRANpkg{pracma} package \citep{pracma_pkg}, which implements
the algorithm proposed by \citet{Higham1988} and produces a positive-definite
approximation under the Frobenius norm.

The released sufficient statistics are calculated after dividing both the
response and predictors by 10. Apart from values affected by clamping, the
corresponding regression therefore has coefficient vector
\((\beta_0/10, \beta_1, \beta_2)\) and residual variance \(\sigma^2/10^2\).
To report the naïve posterior on the original parameter scale, we transform its
mean and covariance using \(A = \operatorname{diag}(10,1,1)\).

\begin{verbatim}
s3_inverse <- pracma::nearest_spd(solve(s3))
parameter_scale <- diag(c(10, 1, 1))

bhat_scaled <- s3_inverse %*% s1
sigma_hat_scaled <- (2 / 10^2) * s3_inverse

bhat <- parameter_scale %*% bhat_scaled
sigma_hat <- parameter_scale %*% sigma_hat_scaled %*% parameter_scale
\end{verbatim}

\begin{figure}[H]

{\centering \includegraphics[alt={Density plots for three regression coefficients compare the privacy-aware analysis in solid red with the naive analysis in dashed red.}]{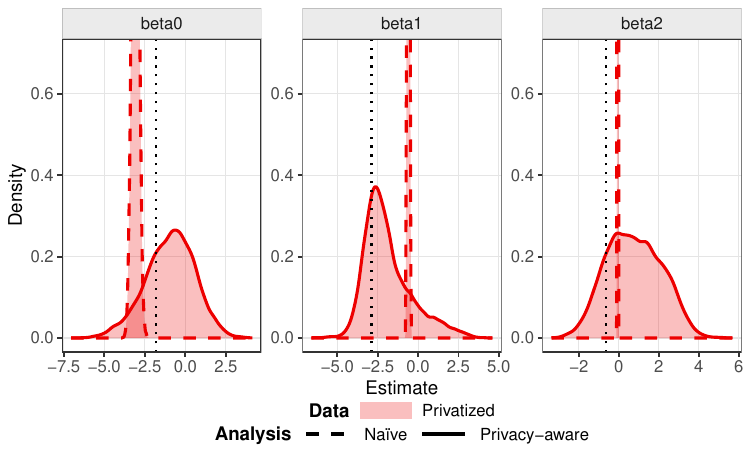} 

}

\caption{(Example 3) Regression-coefficient posteriors based on privatized data
(red). Solid lines denote privacy-aware inference using dapper and dashed lines
denote naïve inference that treats the privatized data as noise-free. Dotted
vertical lines mark the true coefficient values.
Treating the privatized statistics as noise-free produces extremely narrow posteriors displaced from the true coefficient values.}\label{fig:regression-compare}
\end{figure}

\begin{figure}[H]

{\centering \includegraphics[alt={Density plots for three regression coefficients compare confidential-data posteriors in blue with privacy-aware posteriors in red.}]{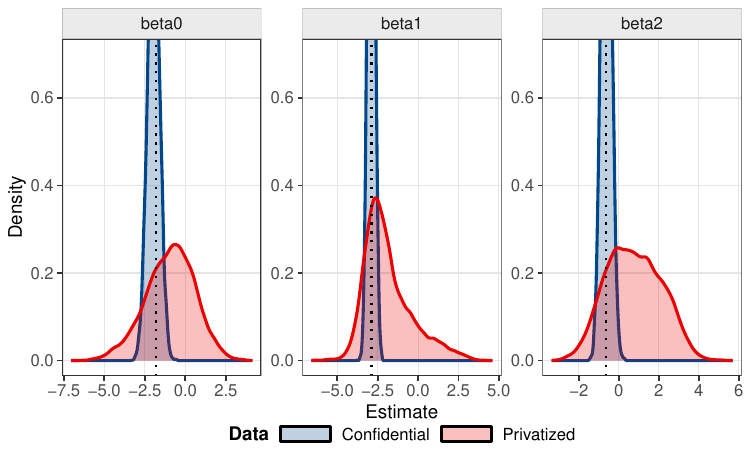} 

}

\caption{(Example 3) Privacy-aware regression-coefficient posteriors using
dapper. Blue denotes confidential data and red denotes privatized data. Dotted
vertical lines mark the true coefficient values.
Accounting for privacy noise yields much broader coefficient posteriors than the confidential-data analysis, reflecting the information lost through privatization.}\label{fig:regression-data-compare}
\end{figure}

Figure \ref{fig:regression-compare} shows the posterior density estimates for the \(\beta\) coefficients based
on \(s_{dp}\). The density estimates indicate the naïve method, which ignores the privacy mechanism, is biased and
underestimates the variance. The especially narrow naïve densities are not a
rescaling artifact. In this simulated release, privacy noise makes the
reconstructed \(x^Tx\) matrix indefinite. The positive-definite approximation to
its inverse nearly collapses one covariance direction, producing spuriously high precision
when the release is treated as noise-free. Applying the same rescaling to the
unnoised statistics recovers the confidential-data covariance in this example.
Likewise, Figure \ref{fig:regression-data-compare}
illustrates how \CRANpkg{dapper} provides point estimates that are not far off from those
that would have been obtained using the original confidential data. The dramatic
increase in the posterior variance indicates the privacy mechanism adds substantial
uncertainty to the estimates.

\subsubsection{Posterior predictive checks}\label{posterior-predictive-checks}

Posterior predictive checks assess whether data generated under the fitted model
resemble the observed data in features relevant to the analysis. In this setting,
the observed data are the privatized sufficient statistics rather than the
confidential records. We therefore reproduce the full data-release process for
each posterior draw: generate a replicated confidential data set with
\texttt{latent\_f()}, calculate its clamped sufficient statistic with \texttt{statistic\_f()},
and add fresh noise from the same Laplace mechanism. This produces draws from
the posterior predictive distribution of the private release.

\begin{figure}[H]

{\centering \includegraphics[alt={Histogram of 500 replicated privatized response sums of squares, with the observed private release marked by a red vertical line.}]{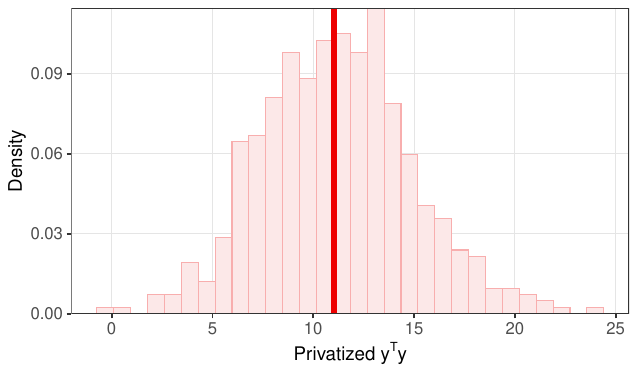} 

}

\caption{(Example 3) Posterior predictive distribution of the privatized response sum of squares. The red vertical line marks the observed private release, and the histogram shows 500 replicated private releases generated under the fitted model. This example demonstrates how dapper can be combined with existing tools such as bayesplot to perform posterior predictive checks on privatized data.}\label{fig:ppc-yty}
\end{figure}

Because \(s_{dp}\) is multivariate, we illustrate the procedure using its \(y^Ty\)
component, which measures the overall magnitude of the response. For every
replicated data set, the fourth component returned by \texttt{statistic\_f()} is the
clamped response sum of squares. Figure \ref{fig:ppc-yty} compares the observed
private value with 500 replicated private values using the \texttt{ppc\_stat()} function
from \CRANpkg{bayesplot}.

\begin{verbatim}
set.seed(123)
ppc_draw_ids <- sample(seq_len(nrow(dp_out$chain)), size = 500)

replicated_private_yty <- vapply(ppc_draw_ids, function(draw_id) {
  simulate_private_yty(as.numeric(dp_out$chain[draw_id, ]))
}, numeric(1))

bayesplot::ppc_stat(
  y = sdp[4],
  yrep = matrix(replicated_private_yty, ncol = 1),
  stat = function(z) z[1],
  bins = 30,
  freq = FALSE
) +
  labs(x = expression("Privatized " * y^T * y), y = "Density") +
  paper_theme +
  bayesplot::legend_none()
\end{verbatim}

This check is limited to one component to demonstrate the
feasibility of posterior predictive checking and is not meant to be a comprehensive
assessment of the multivariate release. A more thorough check can be carried out
by applying the same procedure to the components of \(X^Ty\) and \(X^TX\), or extended using a joint discrepancy
measure. Additionally, prior predictive checks are available by replacing the posterior draws
with draws from the prior, while convergence and posterior diagnostics can be
performed with \CRANpkg{posterior} and \CRANpkg{bayesplot}.

Other parts of the Bayesian workflow can also be easily incorporated.
For instance, simulation-based calibration may be useful for detecting implementation errors in this
setting. One can repeatedly draw parameters from the prior, generate confidential
data and a private release using the user-supplied components, fit the model with
\texttt{dapper\_sample()}, and check whether the generating parameter values have uniform
ranks among the corresponding posterior draws. Systematic departures from
uniformity can reveal inconsistencies among the latent model, statistic, privacy
mechanism, and sampler. Checks requiring an unreleased feature of the confidential data
would require an additional privacy-preserving release.

\subsection{Example 4: wrapper-based sampler}\label{example-4-wrapper-based-sampler}

In all the examples so far, we have used a conjugate prior for the model
of the confidential data. However, many models in practice are more complex and
require MCMC methods to draw samples. In this example, we illustrate how
to use \CRANpkg{dapper} with a non-conjugate prior by
modifying the previous linear regression example to use the \(t\)-distribution
as a prior for the regression coefficients. For this model, we first use the \CRANpkg{fmcmc}
package to construct a sampler for the latent model and then use \CRANpkg{dapper} as a
wrapper around the \CRANpkg{fmcmc} sampler. Because \CRANpkg{dapper} is
modular, the only thing we need to change is the \texttt{posterior\_f()} function. First,
we construct a posterior sampler for the latent data model.\footnote{The prior
  \(t\)-distribution parameters used in this example are for
  illustrative purposes. For a comprehensive discussion of how to
  construct \(t\)-distribution priors for generalized linear models, see
  \citep{Gelman2008}.}

\begin{verbatim}
library(fmcmc)

log_latent_posterior <- function(theta, x, y) {
    #t-distribution as prior
    log_prior_theta <- sum(dt(theta, df = 5, log = TRUE))
    
    mu_y <- cbind(1, x) %*% theta
    log_lik_y <- sum(dnorm(y, mean = mu_y, sd = sqrt(2), log = TRUE))
    
    log_prior_theta + log_lik_y
}
\end{verbatim}

The \CRANpkg{fmcmc} package comes with several proposal distributions. Here
we choose the normal distribution, and some testing shows setting
the scale parameter to 1/5 works well. The sampler returns a \CRANpkg{coda} \citep{coda_pkg}
object that works similarly to objects from the \CRANpkg{posterior} package.
Below is a summary of 2,000 retained post-burn-in draws from our
previous naïve analysis using a \(t\)-distribution with 5 degrees of freedom as
the prior for the regression coefficients. The sampler performs 3,000 total
steps and discards the first 1,000 as burn-in.

\begin{verbatim}
latent_posterior <- MCMC(
    initial   = rep(0, 3), 
    fun       = log_latent_posterior, 
    progress = FALSE,
    nsteps    = 3000, 
    burnin    = 1000,
    kernel    = kernel_normal(scale = 0.2),
    x = dmat[,-1],
    y = dmat[,1] 
  )

summary(latent_posterior)
\end{verbatim}

\begin{verbatim}
#> 
#> Iterations = 1001:3000
#> Thinning interval = 1 
#> Number of chains = 1 
#> Sample size per chain = 2000 
#> 
#> 1. Empirical mean and standard deviation for each variable,
#>    plus standard error of the mean:
#> 
#>        Mean     SD Naive SE Time-series SE
#> par1 -1.849 0.3951 0.008834        0.05980
#> par2 -2.893 0.2125 0.004753        0.01894
#> par3 -0.528 0.2399 0.005363        0.02909
#> 
#> 2. Quantiles for each variable:
#> 
#>         2.5%     25%     50%     75%    97.5%
#> par1 -2.5508 -2.1163 -1.8503 -1.5843 -0.95429
#> par2 -3.2888 -3.0459 -2.8938 -2.7468 -2.44055
#> par3 -0.9773 -0.6817 -0.5338 -0.3822 -0.01746
\end{verbatim}

Next, we use \CRANpkg{dapper} as a wrapper around this sampler to account for
the privacy mechanism. Per-call burn-in can be expensive because it is repeated
at every outer iteration, so it should generally be omitted or kept short when
the wrapped sampler permits. Some adaptive samplers may nevertheless require a
short initialization period. A single valid transition is sufficient for
correctness, but additional transitions may improve mixing. For illustration,
the wrapped sampler below uses \texttt{burnin\ =\ 0} and \texttt{nsteps\ =\ 2}. Because
\CRANpkg{fmcmc} counts the supplied initial state as its first stored step, this
call performs one Markov transition and returns the second state during each
outer \CRANpkg{dapper} iteration.

\begin{verbatim}
posterior_f <- function(dmat, theta) {
    one_step <- MCMC(
        initial   = theta, 
        fun       = log_latent_posterior, 
        nsteps    = 2, 
        burnin    = 0,
        kernel    = kernel_normal(scale = 0.2),
        progress = FALSE,
        x = dmat[,-1],
        y = dmat[,1]
        )
    
    one_step[2,]
}
\end{verbatim}

Finally, to get \CRANpkg{dapper} working with the new sampler,
all we need to do is swap out the definition for \texttt{posterior\_f()} in the privacy model object
and rerun our code. We run one outer chain for 25,000 total MCMC iterations,
discard the first 1,000 as warmup, and retain 24,000 post-warmup draws under the
new \(t\)-distribution prior. A summary of these draws is given below.

\begin{verbatim}
#> # A tibble: 3 x 10
#>   variable   mean median    sd   mad     q5   q95  rhat ess_bulk ess_tail
#>   <chr>     <dbl>  <dbl> <dbl> <dbl>  <dbl> <dbl> <dbl>    <dbl>    <dbl>
#> 1 beta0    -0.765 -0.551  1.40  1.13 -3.36  1.09   1.05     37.0     25.2
#> 2 beta1    -1.68  -1.88   1.46  1.49 -3.66  0.749  1.00     31.4     35.3
#> 3 beta2     0.951  0.932  1.06  1.17 -0.686 2.60   1.07     41.0     88.7
\end{verbatim}

\section{Computational considerations}\label{sec:computation}

The computational performance of \CRANpkg{dapper} depends on both the cost of
each MCMC iteration and the number of iterations required to obtain reliable
posterior estimates. We first report indicative runtimes and effective sample
sizes for the examples in this article. We then discuss how the privacy loss
budget affects mixing and provide practical guidance for diagnosing inefficient
chains.

\subsection{Empirical performance of the examples}\label{sec:benchmarks}

Table \ref{tab:benchmark-results} reports single-run wall-clock times for the
configurations used in the examples. The measurements were obtained on an
eight-core Apple M1 Pro using R 4.6.1 \citep{R_core} and \CRANpkg{dapper} 1.1.1.
The randomized-response example uses
four chains on four parallel workers to demonstrate parallel execution and
multi-chain diagnostics. The remaining examples use one chain and one worker;
these settings were selected in part to keep the demonstrations reasonably
similar in wall-clock duration.

\begin{table}[!h]
\centering
\caption{\label{tab:benchmark-results}Indicative wall-clock runtimes and minimum bulk effective sample sizes for the article examples. Minimum bulk ESS is taken across the model parameters. Under the configurations shown, all four examples finish in under 70 seconds, but their minimum bulk effective sample sizes vary substantially.}
\centering
\resizebox{\ifdim\width>\linewidth\linewidth\else\width\fi}{!}{
\fontsize{7}{9}\selectfont
\begin{tabular}[t]{lrrrrrr}
\toprule
Example & Records & Chains & Iterations (warmup) & Retained draws & Elapsed (s) & Min. bulk ESS\\
\midrule
\cellcolor{gray!10}{Randomized response} & \cellcolor{gray!10}{400} & \cellcolor{gray!10}{4} & \cellcolor{gray!10}{6,000 (1,000)} & \cellcolor{gray!10}{20,000} & \cellcolor{gray!10}{17.1} & \cellcolor{gray!10}{282.2}\\
Discrete Gaussian & 400 & 1 & 2,000 (1,000) & 1,000 & 68.9 & 450.2\\
\cellcolor{gray!10}{Conjugate regression} & \cellcolor{gray!10}{50} & \cellcolor{gray!10}{1} & \cellcolor{gray!10}{25,000 (1,000)} & \cellcolor{gray!10}{24,000} & \cellcolor{gray!10}{62.2} & \cellcolor{gray!10}{133.3}\\
Wrapper regression & 50 & 1 & 25,000 (1,000) & 24,000 & 69.3 & 31.4\\
\bottomrule
\end{tabular}}
\end{table}

The wrapper regression is included to show the overhead associated with calling
an existing MCMC implementation from within \CRANpkg{dapper}. Its inner sampler
performs only one transition per outer iteration, so the increased time when compared
to the conjugate implementation likely
comes from function calling overhead.
More generally, one \CRANpkg{dapper} iteration consists of a posterior update, simulation of a
candidate latent data set, and a record-level Metropolis update for each of the
\(n\) latent records. Consequently, total work is approximately linear in
\texttt{niter}, the number of records, and the number of chains, although independent
chains can be distributed across workers. The constants depend on the cost of
the user-supplied \texttt{posterior\_f()}, \texttt{latent\_f()}, \texttt{statistic\_f()}, and
\texttt{mechanism\_f()} functions; in particular, wrapping a more expensive posterior
sampler increases the cost of every outer iteration.

The single inner transition used in Example 4 is sufficient for correctness,
but increasing the number of inner transitions per call may improve mixing.
This may be especially worthwhile when updating \(\theta\) conditional on the
latent data is inexpensive relative to updating the latent data. The number of
inner transitions is therefore a computational setting that can be chosen using
pilot runs and comparisons of effective sample size per unit of computation.

In practice, users should begin with a shorter pilot run and use
standard MCMC diagnostic tools such as trace plots,
\(\widehat{R}\), bulk and tail ESS, Monte Carlo standard errors, and acceptance
rates to determine whether additional iterations are needed. Multiple chains
from dispersed initial values are preferable when resources permit, and can be
run in parallel using \CRANpkg{furrr}. The outer \texttt{warmup} period should be long
enough to remove initialization effects, while inner burn-in within \texttt{posterior\_f()}
should generally be omitted or kept short because of the cost incurred during every outer iteration.

\subsection{Mixing and privacy loss budget}\label{sec:mixing}

Mixing can be poor when the posterior under a given privacy mechanism is much
more dispersed than the posterior that would arise using the confidential data. In
other words, a small privacy loss budget can result in poor mixing. Intuitively,
this issue arises because the mixing of our Gibbs sampler is constrained by the
coupling between the parameter and the confidential data. Decreasing the privacy
budget will increase posterior uncertainty, until eventually the chain step
sizes become too small to effectively explore the private posterior.
The rest of this section explores a toy example that will provide insight into this
phenomenon.

Suppose the confidential data set consists of a single observation \(x \in \mathbb{R}\),
and consider the scenario where a user makes a request to view \(x\) and
in return receives \(s := x + \nu\), which is a noise-infused version of \(x\).
For simplicity, we do not worry about constructing an \(\epsilon\)-DP mechanism,
and simply take \(\nu \sim N(0, \epsilon^{-2})\) for some \(\epsilon > 0\). However, it
will still be useful to think of \(\epsilon\) as the privacy loss budget, since smaller values
of \(\epsilon\) correspond to a larger amount of noise. Using
a flat prior and a normally distributed likelihood results in
a normally distributed posterior described below.

\begin{equation*}
\begin{aligned}
f(\theta) &\propto 1\\
s \mid x &\sim N(x, \epsilon^{-2})\\
x \mid \theta &\sim N(\theta, \sigma^2)\\
\end{aligned}
\end{equation*}

With the above model, the data augmentation MCMC process consists of the following
two steps:

\begin{itemize}
\item
  Step 1: Sample from \(x \mid \theta, s \sim N(\mu, \tau^2)\), where
  \(\mu\) and \(\tau^2\) are given by
  \begin{equation*}
\begin{aligned}
\mu &:= \dfrac{s/\epsilon^{-2} + \theta/\sigma^2}{1/\epsilon^{-2} + 1/\sigma^2}\\
\tau^2 &:= \dfrac{1}{1/\epsilon^{-2} + 1/\sigma^2}.
\end{aligned}
\end{equation*}
\item
  Step 2: Sample from \(\theta \mid x, s  \sim N(x, \sigma^2)\).
\end{itemize}

In the setting of this example, \citet{Liu1999} showed the Bayesian fraction of missing information
gives the exact convergence rate. The Bayesian fraction of missing information, \(\gamma\), is
defined as
\begin{equation*}
\gamma := 1 - \dfrac{E[Var(\theta \mid s, x) \mid s]}{Var(\theta \mid s)} = 1 - \dfrac{E[Var(\theta \mid x)]}{Var(\theta \mid s)}.
\end{equation*}
Plugging in the appropriate quantities into the above panel gives us
\begin{equation*}
\gamma = 1 - \dfrac{\sigma^2}{\sigma^2 + \epsilon^{-2}} = 1 - \dfrac{1}{1 + \epsilon^{-2}/\sigma^2}.
\end{equation*}
The chain converges faster as \(\gamma \to 0\) and slower as \(\gamma \to 1\).
From the right-hand term in the above panel, we can see \(\gamma\)
depends only on \(\epsilon^{-2}/\sigma^2\), and as the privacy loss budget decreases (i.e.~more noise is being added to \(x\)),
\(\gamma \to 1\).

To illustrate this computational effect in a complete example, we repeated
the conjugate regression analysis from Example 3 over five privacy loss
budgets. We hold the confidential data and a single unit-scale
Laplace noise draw fixed. This single draw is used to produce the
private release for all budgets. Using a common noise source allows us to reduce
extraneous variation between the runs. Each run uses four chains, four
workers, 6,000 total iterations per chain, and 1,000 warmup iterations per
chain, leaving 5,000 retained post-warmup draws per chain and 20,000 pooled
draws for each privacy budget. The four
chains are used here so that \(\widehat{R}\) can be calculated. The original
example uses one chain because its purpose is primarily illustrative.

\begin{table}[!h]
\centering
\caption{\label{tab:privacy-budget-results}Conjugate-regression mixing diagnostics across Laplace privacy loss budgets using coupled noise draws and a fixed MCMC configuration. Minimum bulk and tail ESS and maximum R-hat are taken across the model parameters. This experiment shows that the privacy loss budget can substantially affect mixing.}
\centering
\resizebox{\ifdim\width>\linewidth\linewidth\else\width\fi}{!}{
\fontsize{7}{9}\selectfont
\begin{tabular}[t]{rrrrrrr}
\toprule
Total epsilon & Laplace scale & Elapsed (s) & Min. bulk ESS & Min. tail ESS & Bulk ESS / s & Max. R-hat\\
\midrule
\cellcolor{gray!10}{2.5} & \cellcolor{gray!10}{6.000} & \cellcolor{gray!10}{15.8} & \cellcolor{gray!10}{74.4} & \cellcolor{gray!10}{182.2} & \cellcolor{gray!10}{4.7} & \cellcolor{gray!10}{1.056}\\
5 & 3.000 & 15.4 & 122.3 & 268.5 & 7.9 & 1.035\\
\cellcolor{gray!10}{10 (Example 3)} & \cellcolor{gray!10}{1.500} & \cellcolor{gray!10}{15.3} & \cellcolor{gray!10}{163.7} & \cellcolor{gray!10}{274.7} & \cellcolor{gray!10}{10.7} & \cellcolor{gray!10}{1.018}\\
20 & 0.750 & 15.2 & 240.8 & 479.6 & 15.9 & 1.011\\
\cellcolor{gray!10}{40} & \cellcolor{gray!10}{0.375} & \cellcolor{gray!10}{15.2} & \cellcolor{gray!10}{417.9} & \cellcolor{gray!10}{743.2} & \cellcolor{gray!10}{27.5} & \cellcolor{gray!10}{1.008}\\
\bottomrule
\end{tabular}}
\end{table}

As Table \ref{tab:privacy-budget-results} shows, elapsed time remains nearly
constant across these runs, whereas the number of effectively independent
draws changes substantially. Over the range considered, the minimum bulk ESS
increases from 74.4 to 417.9 and the maximum \(\widehat{R}\) decreases from
1.056 to 1.008 as the privacy loss budget increases from 2.5 to 40. The
value used in Example 3, \(\epsilon=10\), yields a minimum bulk ESS of 163.7 and a maximum
\(\widehat{R}\) of 1.018. These diagnostic patterns are consistent with better mixing
at larger budgets.

This comparison is restricted to one privacy mechanism and one
model, so it should be interpreted with caution.
Although \(\epsilon\) has a common interpretation as a worst-case privacy-loss bound, its numerical value is not a global scale for MCMC performance. At the same \(\epsilon\), a sampler may mix well under one privacy mechanism but poorly under another.
As such, privacy-budget sensitivity checks should be interpreted within the
specific mechanism, statistic, and model being analyzed.

If one cannot achieve good mixing with acceptable compute
times, we recommend artificially varying the privacy
budget to see if mixing significantly improves.
If this is the case, then \CRANpkg{dapper} may be poorly suited for the problem at hand,
and alternative methods, such as the pseudo-marginal method \citep{Andrieu2009},
can be pursued.
Additionally, since programming samplers can be a bug-prone endeavor, it may be helpful
to artificially vary the privacy loss budget as a
sanity check; slow mixing could be due to an
error introduced when specifying the components of \CRANpkg{dapper}. For example, one can
generate synthetic data using \texttt{latent\_f()} and compare the posterior
draws from \texttt{posterior\_f()} and \texttt{dapper\_sample()}. Generally speaking,
results should look similar when using a sufficiently large privacy loss budget.

\section{Summary}\label{sec:summary}

Differential privacy powers many state-of-the-art data privacy
implementations. Software supporting DP deployment and analysis
is growing rapidly, with several large companies building software ecosystems for
working with differential privacy
(e.g.~AWS Clean Rooms \citep{awscleanrooms2024}, SmartNoise \citep{smartnoise2024}, and PipelineDP \citep{pipelinedp2022}).
However, the majority of these software tools only address privacy and not the
ensuing analysis or, if they do, only address the analysis for specific models.
Furthermore, none of the mentioned software suites are well suited for R users.

\CRANpkg{dapper} helps fill a capability gap by providing researchers a
way to perform privacy-aware Bayesian inference in R. As summarized in Table
\ref{tab:software-comparison}, most related tools construct private releases
or fit models directly to confidential data. \CRANpkg{SimBaRepro} is an
important exception that also supports inference from already privatized data,
but provides frequentist tests and confidence sets rather than posterior draws
or a point estimate.

This package offers a significant step forward in extending general-purpose statistical inference tools from confidential to privatized data.
Despite its strengths, there are
several cumbersome requirements for good performance that limit its potential.
First, the privacy mechanism's log density or log probability mass must be evaluable.
Second, a record-additive
statistic must be used to leverage \CRANpkg{dapper}'s full computational potential.
Third, the non-private posterior sampler needs to mix well. Finally, a very small privacy loss budget can result in excessively slow mixing.
Future versions of \CRANpkg{dapper} could incorporate methodological developments that relax some of these requirements.

\section*{Supporting software}\label{supporting-software}
\addcontentsline{toc}{section}{Supporting software}

Data preparation uses \CRANpkg{tidyverse} \citep{tidyverse_pkg}, including
\CRANpkg{dplyr} \citep{dplyr_pkg}, \CRANpkg{tibble} \citep{tibble_pkg}, and
\CRANpkg{tidyr} \citep{tidyr_pkg}. Figures use \CRANpkg{ggplot2} \citep{ggplot2_pkg},
\CRANpkg{ggsci} \citep{ggsci_pkg}, and \CRANpkg{scales} \citep{scales_pkg}, alongside
the diagnostic plotting tools described above. The examples also use
\CRANpkg{MASS} \citep{MASS_pkg} for multivariate normal simulation and
\CRANpkg{VGAM} \citep{VGAM_pkg} for Laplace densities and simulation. Parallel
execution uses the \CRANpkg{future} framework \citep{future_pkg} through
\CRANpkg{furrr}.

The article is generated with \CRANpkg{rmarkdown} \citep{rmarkdown_pkg},
\CRANpkg{knitr} \citep{knitr_pkg}, and \CRANpkg{rjtools} \citep{rjtools_pkg}, with
\CRANpkg{kableExtra} \citep{kableExtra_pkg} for table formatting.

\section*{Disclosure statement}\label{disclosure-statement}
\addcontentsline{toc}{section}{Disclosure statement}

Kevin Eng and Ruobin Gong are grateful for support from the Alfred P. Sloan Foundation. Jordan Awan's research was supported in part by NSF grants SES-2150615 and SES-2610910. NJ acknowledges support from the NSF grant DMS-2610236. OpenAI's Codex was used for light editing and assistance with selected tables and figures. All outputs were reviewed and verified by the authors.

\bibliography{RJreferences.bib}

\address{%
Kevin Eng\\
Rutgers University\\%
Department of Statistics\\ Piscataway, NJ 08854\\
\href{mailto:ke157@stat.rutgers.edu}{\nolinkurl{ke157@stat.rutgers.edu}}%
}

\address{%
Jordan A. Awan\\
University of Pittsburgh\\%
Department of Statistics\\ Pittsburgh, PA 15260\\
\url{https://jordan-awan.com/}\\%
\href{mailto:jaa557@pitt.edu}{\nolinkurl{jaa557@pitt.edu}}%
}

\address{%
Nianqiao Phyllis Ju\\
Dartmouth\\%
Department of Mathematics\\ Hanover, NH 03755\\
\url{https://nianqiaoju.github.io/}\\%
\href{mailto:nianqiao.ju@dartmouth.edu}{\nolinkurl{nianqiao.ju@dartmouth.edu}}%
}

\address{%
Vinayak A. Rao\\
Purdue University\\%
Department of Statistics\\ West Lafayette, IN 47907\\
\url{https://varao.github.io/}\\%
\href{mailto:varao@purdue.edu}{\nolinkurl{varao@purdue.edu}}%
}

\address{%
Ruobin Gong\\
Rutgers University\\%
Department of Statistics\\ Piscataway, NJ 08854\\
\url{https://ruobingong.github.io/}\\%
\href{mailto:ruobin.gong@rutgers.edu}{\nolinkurl{ruobin.gong@rutgers.edu}}%
}

\end{article}

\end{document}